\documentclass[11pt]{article}
\usepackage{geometry}                
\usepackage{graphicx}
\usepackage{epstopdf}
\usepackage{amsmath}
\usepackage{amssymb}
\usepackage{color}
\usepackage{colordvi}
\usepackage{colortbl}
\usepackage{placeins}

\usepackage{setspace}
\usepackage{url}
\usepackage{adjustbox}

\usepackage{mdframed} 
\usepackage{thmtools}

\usepackage{makeidx}
\makeindex

\newcommand{\qed}{\hfill $\Box$}

\newcommand {\ul}[1]      {\underline{#1}}

\newcommand{\komment}[1]{}

\definecolor{LightGreen}{rgb}{0.8,1,0.8}
\definecolor{LightRed}{rgb}{1,0.8,0.8}
\definecolor{LightGrey}{rgb}{0.8,0.8,0.8}
\definecolor{NotSoLightGrey}{rgb}{0.7,0.7,0.7}
\definecolor{MyGray}{gray}{0.9}

\newcounter{xexersicecounter}[section]

\newtheorem{prop}{Proposition}[section]

\newtheorem{theorem}[prop]{Theorem}

\definecolor{TUMhellblau}{rgb}{0.596,0.776,0.918}
\definecolor{TUMblau}{rgb}{000,0.396,0.741}

\newcommand{\boxit}[1]{\fbox{\begin{minipage{\textwidth}#1\end{minipage}}}}

\title{Filter Bubbles, Echo Chambers, and Reinforcement: Tracing Populism in Election Data}

\author{Johannes M\"uller\footnote{Center for Mathematics, Technische Universit\"at M\"unchen, 85748 Garching, Germany} 
	\footnote{Institute for Computational Biology, Helmholtz Center Munich, 85764 Neuherberg, Germany}, \ 
Volker H\"osel$^\ast$
Aur\'elien Tellier\footnote{Section of Population Genetics, Center of Life and Food Sciences Weihenstephan, Technische Universit\"at M\"unchen, 85354 Freising, Germany}} 

\begin{document}
	\maketitle

\begin{abstract}
	We present a novel model for the effect of echo chambers, filter bubbles, and reinforcement on election results. Our model extends the well known voter model with zealots to include reinforcement. We analyze the behaviour of the model, determine the invariant measure, and show that reinforcement may 1) shift the distribution of votes compared to the voter model, and 2) lead to phase transitions. 
	We test whether the model with reinforcement fit better than under its absence in election data from US presidential elections, the Brexit referendum, and parliamentary elections in France, The Netherlands, and Germany. We find in many cases that populist parties and candidates can be clearly identified by a high level of reinforcement. Furthermore, we find the phase transition predicted by the model back in data. We finally discuss the implications and relevance of our findings and possible origins of the reinforcement behaviour in modern societies.
\end{abstract}

\section{Introduction}

We are living in the age of rising populism. Such statement have become abundant in recent years \cite{GuardianPopulism}, while there is no unique, commonly accepted definition of the term ``populism'', in the political and sociological literature. Different authors use a wide variety of concepts to narrow down the term ``populism''~\cite{Gidron2013} making use of economic, social, and political aspects, as well as  organization, communication, or the personality of the centrally involved figures. Among others, Mudde \cite{Mudde2004} and Mudde and Kaltwasser \cite{Mudde:Book} developed the idea of populism as a ``thin-centered ideology''. In that fruitful interpretation, populism expresses a structural and organizational framework that can be associated with any ideology rather than being an ideology in itself. One central aspect proposed by Mudde \cite{Mudde2004} is the segregating habit of populist movements, dividing the citizens into two opposing groups, a (corrupt, evil) elite and the (ordinary, plain and good) people. If a nationalistic ideology is involved, the term ``people'' also incorporates the notion of national community in contrast to immigrants (even when it is historically and sociologically unjustifiable). Populists divide the population in a Manichean outlook into an in- and an out-group, where the in-group is good and accepted, while the out-group is bad and must either be converted or fought. 
\par\medskip 
This segregating mindset of populists affects communication structures, that is the conception, release and acceptance of news and fake-news. Members of a populist group tend in a higher degree than the average population to accept uncritically the opinion of their own group and to refuse other opinions~\cite{Arceneaux2012}. The opinion, or even fake-news, is not literally accepted but may rather be a sign of group membership. This aspect of the populist communication structure might not be predominant in all cases, but seems to be a logical consequence of the populists' world understanding. Populist parties therefore share a common rhetoric (see the investigation by the Populism Research Team, \cite{HawkinsPopulism}) which fuels reinforcement, filter bubbles, and echo chambers. According to this hypothesis, the communication by populist parties reinforce the belief of their partisans (or voters) and in the same time the partisans isolate themselves in a filter bubble in which only certain types of information can penetrate. A group of populist partisans would thus constitute an echo chamber in which certain points of view, while being dismissed as low value or fake in the general population, become amplified and accepted as such. We focus in the present study on that aspect of reinforcement, filter bubble and echo chamber. Our aim is here to develop a variant of the well known mathematical voter model to include reinforcement, and then to perform statistical analysis of election results to reveal traces of reinforcement. In recent years, statistics and modeling approaches investigating the appearance 
and effects of filter bubbles have been developed~\cite{Geschke2018,Bail2018,Flaxman2016}. 
However, this study represents to our knowledge the first attempt to reveal the signatures of reinforcement directly in election data. \par\medskip 

Statistical patterns in election data are observed and investigated since several years~\cite{CostaFilho1999,Palombi2015}. This novel investigation into election results is built on the principles that mathematical models can predict expected statistical patterns which can be tracked down in the data. By deriving models with different hypotheses, it is possible to study the underlying mechanisms generating the observed patterns and election results. Mathematical models have a long tradition in political sciences. Two general classes of mechanisms, neutral and non-neutral, can be studied. Non-neutral approaches take the political direction of parties into account. The most prominent representative in this line of research is the Hotelling model~\cite{hotelling1929}, that uses game theory to understand how parties position themselves in the political arena. That model is refined, e.g. into the valence model, which is able to describe the political landscape of many western democracies~\cite{SchofieldIsrael2005,SchofiledBrit2005,SchofieldNLGerm1998}.
A line of thinking that gained importance recently is based on neutral models~\cite{Dong2018}. The idea is that many of the observed statistical features of election results can be explained by models that do not take into account the political direction of parties. In these models parties are assumed to stand for groups of individuals who are committed to a common attitude. These models are particularly used to instigate the turnout problem formulated for the first time by Downs~\cite{Downs1957}: Why should a single voter take the effort and go to vote, if most likely the outcome of the election will not change? A combination of game theory and stochastic models~\cite{Palfrey1983,Myerson1998}, but also agent based simulation models~\cite{Bendor2003} have been developed to address this problem. 
More fundamentally, the noisy voter model~\cite{Fernandez-Gracia2014,Redner2019} addresses communication of voters to understand the outcome of elections. In the context of elections, the noisy voter model is called the zealot model. Zealots or partisans are individuals (or other influencing entities as newspapers) that do not change their political direction but stick to one given group, and influence other population members. 
It is successfully used to explain the variance structure in election data ~\cite{Braha2017,Kononovicius2017}, the fraction of swing voters who readily change their favorite party~\cite{Sano2017}, or to obtain an idea about the time to consensus~\cite{mobilia2013}. 
The voter model with party dynamics is an alternative approach, where the fixed number of parties in the zealot model is replaced by a dynamical mechanism, in which parties can be crated and vanish again~\cite{Hoesel2019}. This latter approach explains the log-linear structure in election data observed in several countries as soon as more than 10 candidates (or parties) are present ~\cite{Hoesel2019}. A noisy-voter model with linear branching process can also generate distributions of candidates' vote shares when the candidates are in the same party ~\cite{Fortunato2007}. Note, however, that mathematically these models do not allow for bifurcations and phase transitions -- similar to disruption, phase transitions lead to sudden fundamental changes in the system though the situation seemed to be only slightly changed. 
Few models exhibiting phase transitions are nevertheless used in mathematical sociology~\cite{Galam2008,Mimkes2006}. Of special relevance, Nicolao and Ostilli used a Potts model to investigate twitter data related to elections~\cite{Nicolao2019}. The Potts model shows a phase transition for certain parameter combinations. Interestingly, the authors identify in the data parameters that often are close to those critical parameter values. In some cases, the data indicated that the twitter system spontaneously breaks the symmetry and can generate different outcome/behavior, which is to expect for a supercritical system ~\cite{Nicolao2019}. \par\medskip 

In the present paper, we augment the zealot model by integrating reinforcement. We thus want to investigate if such neutral model with reinforcement captures observed patterns in election data, and thus if reinforcement, filter bubbles, and echo chambers are important enough for populist parties or candidates to leave statistical signatures in election data. First, we develop the model and investigate its mathematical behavior. By investigating a strong- and a weak-effects limit for large population sizes, we find bifurcations and phase transitions. In the limit of weak effects, we derive explicitly the invariant measure for the vote share for our reinforcement model. Second, we analyze election data from several elections (US presidential, Brexit in the UK, parliamentary in Germany, Netherlands and France) and show that often the reinforcement model describes the data appropriately, and significantly better than the zealot model. We conclude that reinforcement and echo chambers represent forces that leave signatures in the empirical data. Astonishingly, it is often (but not always) possible to clearly identify populist parties and candidates as they show a high reinforcement parameter (US, Germany, Brexit, but not France). Furthermore, the traces of phase transitions are visible in election data (The Netherlands). Finally, we discuss the implications and relevance of our results and limitations of our modeling/statistical approach.

\section{Model, model analysis, and model predictions}

In this section, we modify the zealot/noisy voter model to include reinforcement and echo chambers. In the underlying basic voter model (without zealots)~\cite{Liggett1985}, each individual adopts an opinion (opinion~1 or opinion~2). At rate $\mu>0$, a person rethinks his/her opinion, and copies the opinion of a randomly chosen population member. In the situation of a finite, homogeneously mixing population, this Markov process will end up in an absorbing state, where only one opinion remains. The other opinion dies out. This outcome is naturally unrealistic, and thus the zealot model~\cite{Braha2017} assumes an additional mechanism. Apart of floating voters, i.e. individuals who change their opinion, there are zealots. Zealots are individuals with a strong opinion and thus do not change their mind, even when in contact with a distinct opinion whether from other individuals, newspapers or other mass media. Moreover, zealots influence the opinion of floating voters, as a floating persons copies the opinion of a randomly selected individual out of the group of all individuals (floating voters or zealots). In the zealot model, all opinions persist at all time as the model has no absorbing state and a non-trivial invariant distribution ~\cite{Braha2017}. \\

We model reinforcement and echo chambers as follows. An individual who ``live'' in an echo chamber does not interact with a representative sample of the population, but only with a sub-group of individuals who are more likely to share the same opinion. 
We introduce weights $\vartheta_i\in(0,1]$ to express the reduced interaction with the opponent group. We want to compute the share of votes for each party as a function of the model parameters (especially the reinforcement). \\

Let $N$ denote the total population size, $N_i$ the number of zealots with opinion $i\in\{1,2\}$, and $\vartheta_i\in(0,1]$ the weights for the opposite opinion. If $X_t$ is the number of supporters for opinion~1, while $N-X_t$ is that for opinion~2, then
\begin{eqnarray}
X_t\rightarrow X_t + 1 &\mbox{ at rate } & \frac{ \mu (N-X_t)\,\vartheta_1 (X_t+N_1)}{\vartheta_1(X_t+N_1)+(N-X_t+N_2)},\\
X_t\rightarrow X_t - 1 &\mbox{ at rate } & \frac{ \mu X_t\,\vartheta_2 (N-X_t+N_2)}{(X_t+N_1)+\vartheta_2 (N-X_t+N_2)}.
\end{eqnarray} 
\par\medskip 

Note that $\vartheta_1$ is the probability of group-2-individuals to interact with group~1, and $\vartheta_2$ that of group-1-members to interact with group~2. To be clear, $\vartheta_2$ measures thus the strength of reinforcement of group~1. Obviously, this model agrees with the zealot model in case of $\vartheta_1=\vartheta_2=1$.\\ 
In order to understand the properties of the model, we investigate the vote share in the continuum limit $N\rightarrow\infty$. As it is well known from parallel investigations in population  genetics~\cite{tavare:book}, two different scales of the parameters are sensible: 1) the number of zealots scale linearly with $N$, s.t.\ $N_i=n_i\, N$, while the reinforcement parameters $\vartheta_i$ are constant in $N$ (the deterministic limit) or 2) $N_i$ are constant in $N$ and $\vartheta_i = 1+{\cal O}(N^{-1})$, called the weak effects limit in which the effect of zealots and reinforcement become small if the population size becomes large. \par\medskip 

\begin{figure}[t!]
\begin{center}
	\includegraphics[width=12cm]{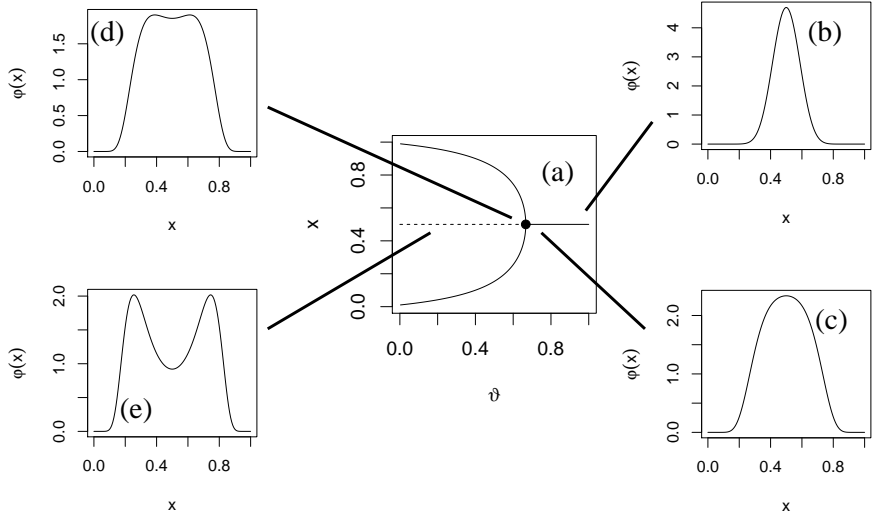}
\end{center}	
\caption{Behavior of the zealot model with reinforcement. (a): Stationary states (solid lines: locally asymptotically stable, dashed line: unstable) and pitchfork bifurcation (indicated by a bullet) of the deterministic limit. Panels (b)-(d): Corresponding  invariant measures of the weak-effects limit. Parameters (a): $n_1=n_2=0.1$, $\vartheta_1=\vartheta_2=\vartheta$; (b)-(d): $N_1=N_2=20$, and 
	(b) $\theta_1=\theta_2=10$, 
	(c) $\theta_1=\theta_2=70$, 
	(d) $\theta_1=\theta_2=80$, 
	(e)~$\theta_1=\theta_2=100$.
}\label{fig1}
\end{figure}

We start with the deterministic limit. If the intrinsic and extrinsic influences are strong, often the state can be well approximated by a normal distribution with a variance that declines with the inverse of the population size. In the limit, we obtain an ordinary differential equation (ODE) for the fraction of opinion-1-supporters, ($n_i$ are the ratio of zealots over free voters, $N_i/N$, for a derivation see the supplementary information, SI)
\begin{eqnarray}\label{reforceODE}
\dot x &=& 
-\mu x\,\frac{\vartheta_2 (1-x+n_2)}{(x+n_1)+\vartheta_2 (1-x+n_2)}\\
&&+
\mu (1-x)\,\frac{\vartheta_1 (x+n_1)}{\vartheta_1(x+n_1)+(1-x+n_2)}.\nonumber
\end{eqnarray}
It is possible to analyze bifurcations, that is change-points in the structure of stationary solutions (see Fig.~\ref{fig1} (a) and SI). \par

Particularly, in the symmetric case (all group-specific parameters $N_i$, $\vartheta_i$ are identical), we find a Pitchfork bifurcation if the reinforcement becomes large. If $\vartheta_i$ (the probability to interact with the opposite group) drops below a certain threshold, suddenly two stationary states appear, both of them being locally asymptotically stable. One of the two opinions does prevail, and the other opinion (which has the potential to also prevail) cannot influence the population strongly enough to change this state. In non-symmetric cases, the group that follows a strong reinforcement strategy also has a strong advantage versus the opponent's group. Even if the opponent group has way more zealots, the opinion with the reinforcement strategy does prevail (see SI).
\par\medskip 

On the one hand, the deterministic limit allows to easily analyze and visualize the long term behavior. On the other hand, population genetics studies ~\cite{tavare:book} show that the weak effect models are more realistic when it comes to the analysis of data. In this case, the zealots are constant in number, and the reinforcement declines with the inverse of the free voters' number, $\vartheta_i=1-\theta_i/N$. Under these assumptions we obtain an explicit formula for the invariant distribution of group 1's vote share $x$ for $N\rightarrow\infty$, (SI)
\begin{eqnarray}
\varphi(x) =  C\,e^{\frac 1 2 (\theta_1+\theta_2)x^2-\theta_1\,x}\,\, x^{N_1-1}\,(1-x)^{N_2-1}.\label{reinfDirichDist}
\end{eqnarray}
The second term,  $x^{N_1-1}\,(1-x)^{N_2-1}$, only depends on the zealot's numbers. This term is a beta distribution, as predicted by the zealot model. The Beta-distribution is modified by the first factor $e^{\frac 1 2 (\theta_1+\theta_2)x^2-\theta_1\,x}$ that reflects the effect of reinforcement. The multiplicative constant $C$ ensures that the integral of the distribution is $1$.\\
As indicated in Fig.~\ref{fig1}~(b)-(d), the bifurcations in the deterministic limit can be found back in the invariant measure of the weak-effects limit. If the scaled reinforcement parameters $\theta_i$ become large, the unimodal distribution becomes bimodal, which corresponds to the pitchfork bifurcation observed above. In terms of statistical physics~\cite{Krapivsky2017}, the positive feedback introduced by the reinforcement mechanism is able to drive the system into a phase transition. 
\par\medskip

\section{Data analysis}

We compare in the following the theoretical predictions with empirical findings. Hereby, we assume that each election district is independent of all other districts, and identical realization of the invariant distribution of the democratic process. For each election, we obtain the results of voting for many election districts and use a maximum-likelihood method to estimate the model parameters (reinforcement parameter for each candidate/party). The Kolmogorov-Smirnov test (KS) is applied to assess the consistency of the theoretical model to empirical data, namely by obtaining the p-value that the distribution of vote share for one party against the sum of the other ones does fit our prediction from the reinforcement or the zealot model. The likelihood-ratio test (LL) to compare the performance of these models. We obtain also an estimate of the reinforcement parameter for each candidate/party.\par\medskip

For the data analysis, we assume that the results in different election districts are independent replicas of the election. This assumptions is, of course, a simplification. First, there is a spatial correlation in the election results~\cite{Borghesi2012}. Second, election districts are influenced by social covariates or nuisance variables as income or the dichotomy rural/city area. Third, we developed our theory for dichotomous elections, but apply the results also to multi-party systems in focusing on one given party, and lumping all other parties together into one pseudo-group. However, if we pool the districts, it is possible that some if the perturbing influences become less important. And indeed, the analysis shows that the data are described quite well. 
\par\medskip

\begin{figure}[h!]
	\begin{center}
        (a) \includegraphics[width=7cm]{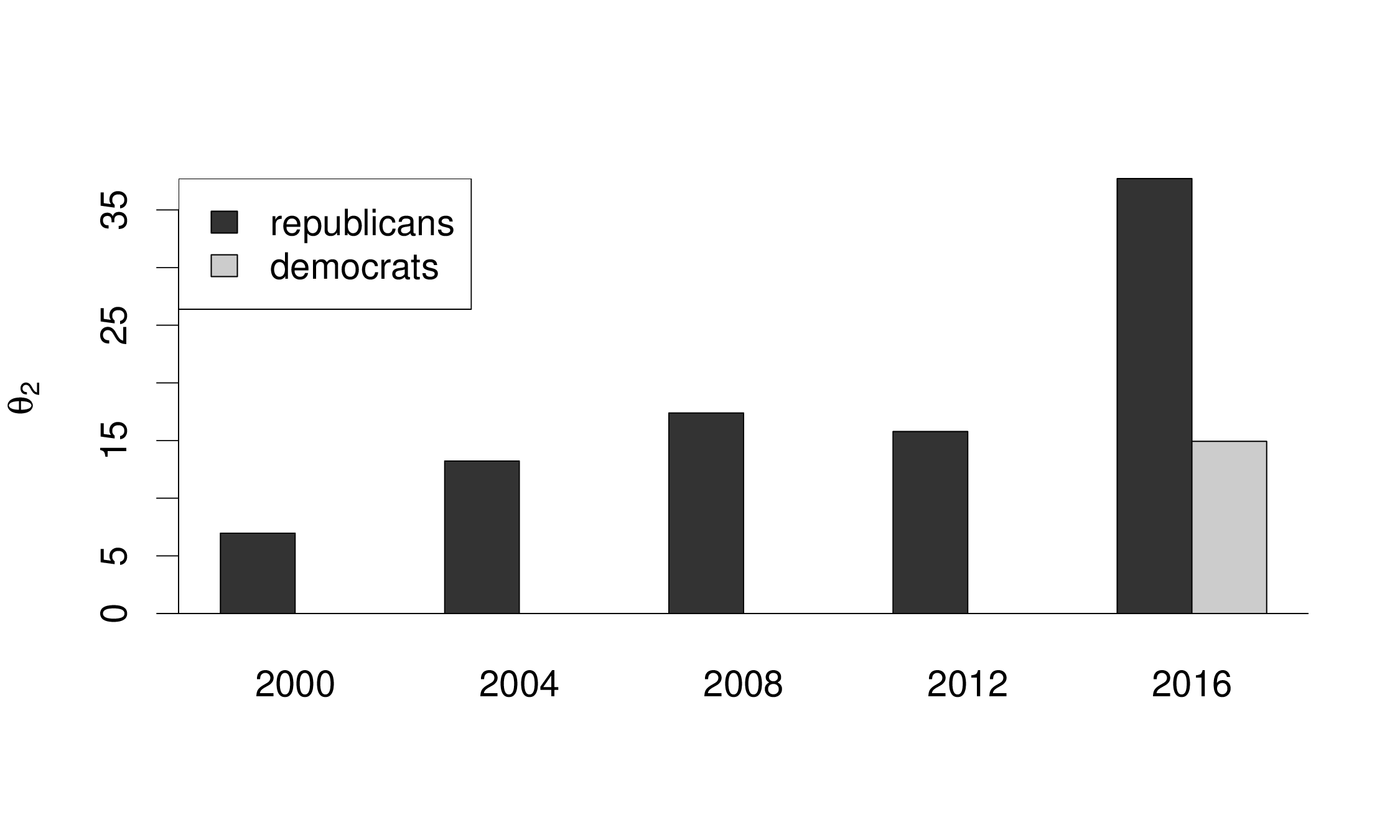}	
		(b) \includegraphics[width=7cm]{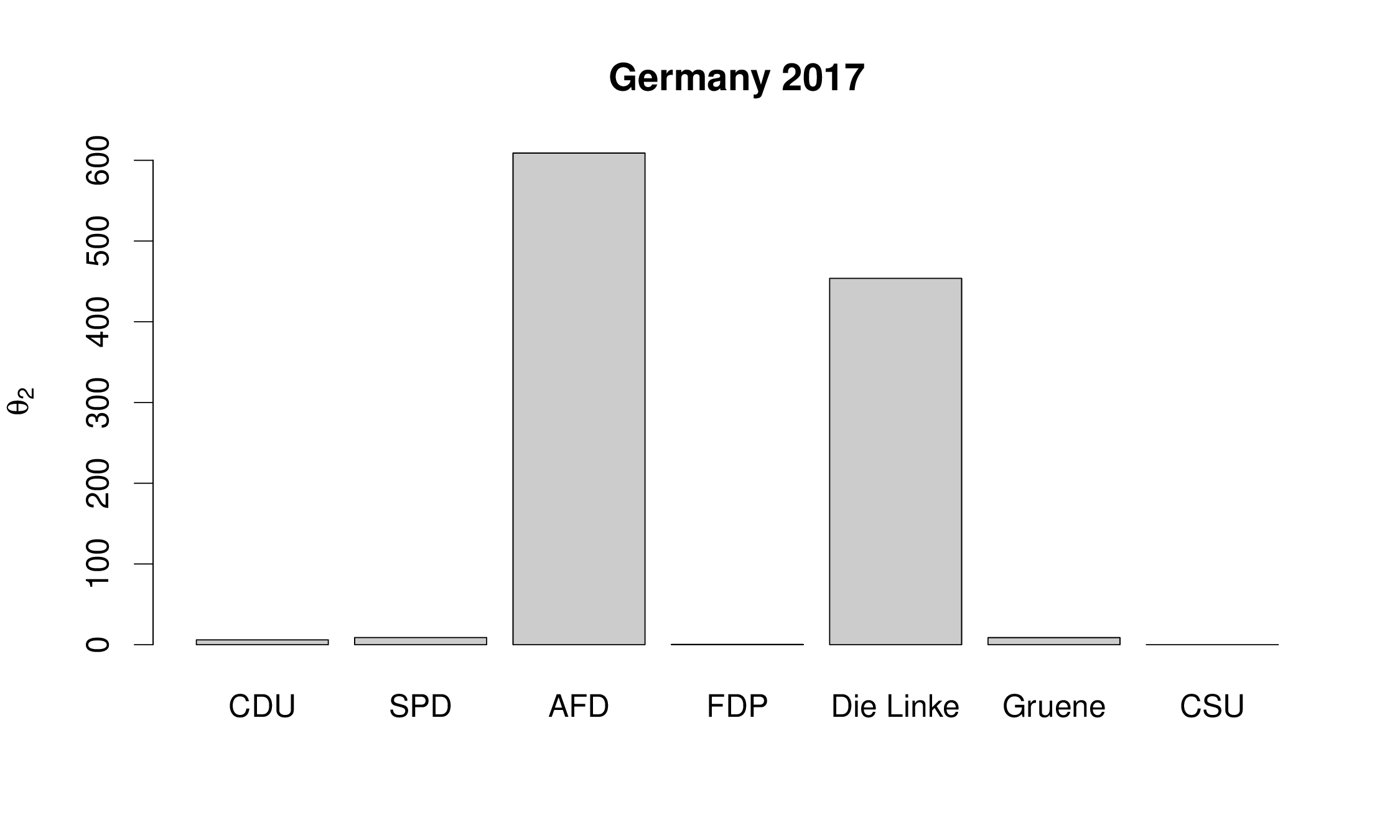}
		\end{center}
	\caption{Estimates of reinforcement parameters for elections in the USA (a), in Germany (b). (a) Estimates of the reinforcement parameter for the USA presidential elections in 2000--2016. Note that in 2000-2012, the reinforcement-parameter for the democrats is negligible small, and only in 2016 it becomes visible. (b) Reinforcement parameter for several parties, from the parliamentary elections in 2017 for whole Germany.}
\label{fig:two}
\end{figure}

\paragraph{US presidential elections} 
We find a good agreement of the reinforcement model and data for  the recent US presidential elections 2000-2016 (Fig.~SI, Fig.~1). The KS test yields $p$-values larger or equal $0.1$ apart of the year 2000. This finding indicates that the reinforcement model cannot be rejected. The year 2000 is kind of exception, as also the green party did win a considerable percentage of votes. For the 2000 and the 2004-elections, the zealot and the reinforcement models fit equally well. For the election results between 2008-2016 the reinforcement model is superior (LL-test), and the zealot model is not appropriate (KS, see SI for details, point estimates and test results).
When inspecting the reinforcement parameters for the two parties (Fig.~\ref{fig:two}), we find that the reinforcement for the democrats is rather unimportant (in comparison to that of the republicans) before 2016. In the 2016 election, the reinforcement of the democrats jumps to the averaged value of the republicans before 2016, while the reinforcement of the republicans shows a more than twofold increase (Fig.~\ref{fig:two}). We associate this effect to the consequences of the populist attitude of the republican candidate Donald Trump. 
\par\medskip 

\paragraph{Brexit}
The reinforcement-model is statistically clearly superior to the zealot model to describe the Brexit data (likelihood-ratio test $p < 10^{-10}$, see also SI, Fig.~2). Also the Kolmogorov-Smirnov test clearly indicates that the reinforcement-model cannot be rejected ($p=0.75$), but the zealot model that does not incorporate reinforcement is not appropriate (KS: $p=  0.0009$). Interestingly, the point estimates for $\theta_i$, suggests that the brexiters are responsible for over 99\% of the reinforcement. 
In order to investigate this finding more thoroughly, we use the likelihood-ratio test to compare a restricted model where both groups do have the same reinforcement parameters ($\theta_1=\theta_2$) and the model where the reinforcement parameter for both groups are arbitrary. It turns out that the model that allows for more reinforcement in brexetiers than in remainers is highly superior (the hypothesis that the reinforcement parameters for both groups are identical is rejects at $p<$1e-10). The model hints to the fact that populist tendencies have been at the brexetiers part, and only in a minor amount in the remainders side.

\paragraph{Germany}
If we investigate the detailed parliamentary (Bundestag) election results from 2017 for the seven parties that are present in the parliament, we find that reinforcement plays a role particularly for two parties: ``AFD'' and ``Die Linke''. Both parties are known as populist 
parties~\cite{Kim2017,Hough2009,Hough2019} on either side of the political spectrum. For all other parties, reinforcement is not statistically significant. We further find that the reinforcement of the AFD is strongly connected with the ``new'' states (the states/regions that are located in the former DDR/Eastern Germany) and not in the ``old'' states (states/regions of the former Western Germany, see SI), while the reinforcement of ``Die Linke'' seems to be independent of the geographic location of ``old'' and ``new'' states. 

\paragraph{France} 
We analyzed the election data for the first and the second round of the presidential elections to detect signatures of reinforcement. When especially studying the 2017 elections, and focusing on the candidate of the Front National, Marine Le Pen,  the reinforcement model does not fit better than the zealot model (LL: $p\approx 1$). These results are obtained at the scale of department administrative units as well as at the smaller units of the canton. From our analysis point of view, it appears that Marine Le Pen does not exhibit the populist reinforcement observed for the republicans in the US, the Brexiters in the UK and the AFD party in Germany. When studying the first round of the presidential elections since 1965, we find that in several instances, the reinforcement model fits better than the zealot model, for example (SI,Table~5) for De Gaulle (1965), Chaband-Delmas (1974), Chirac (1981, 1988, 1995, 2002), Jospin (2002), Sarkozy (2007, 2012), Hollande (2012) and Fillon (2017).  The parameter of reinforcement often is above 0.5 meaning that if there is any effect, it reflects the ability of the candidates to retain their supporters in the long term. It is also possible, that the model -- that is only build for two parties - fails to appropriately interpret the data of the multi-party system in France.
\par\medskip 

\begin{figure}[t!]
	\begin{center}
		 \includegraphics[width=7cm]{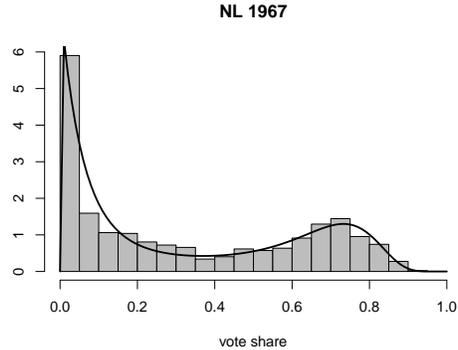}
	\end{center}
	\caption{The model result for the 1967 election in The Netherlands where the focal group is the  Catholic People's Party. We did fit the model using the focal group versus the pooled remaining groups.}
	\label{supercrit}  
\end{figure}

\paragraph{Phase Transition}
Our model predicts that a phase transition is possible which impacts the outcome. In effect, some groups may become supercritical, leading to  a bimodal distribution appearing in the election data. We observe such bimodal distributions in  The Netherlands 1967 parliamentary elections (Fig.~\ref{supercrit}).
\par\medskip    

The Netherlands have a proportional voting system, where many parties compete for votes. In order to handle this situation, we focus on one group and merge all other groups in a pseudo-party. As shown in Fig.~\ref{supercrit}, the model is able to fit the overall structure of the data. The Dutch party we address is the "Katholieke Volkspartij (KVP)", the Catholic People's Party. After the second world war, this party played a major role in The Netherlands with a vote share of about 0.33; only after 1967, the vote share dropped to around 0.25. During the active time of the party, especially before 1971, the reinforcement model is highly superior to the zealot model (LL: $p<10^{-10}$). We clearly find a bimodal distribution (Fig.~\ref{supercrit}). \par\medskip

\section{Discussion}

A first major result of our study is that neutral models, that is, models that do not take political content into account, are able to reveal fundamental structures in the political process. At present, such models do not play a major role in the scientific discussion. The situation resembles that of population genetics and ecology 20 years ago, when neutral models for the analysis of fundamental evolutionary and ecological mechanisms have been proposed. Only after a long discussion (that partially is still going on), these tools have been accepted as a valuable way to access important research questions~\cite{Clark2009,Kern2018,Jensen2019}. We expect that also in the political sciences, models as those proposed in the present study gain importance.
\par\medskip 
The second central finding of the present paper is that the method proposed is able to detect reinforcement in election data. Interestingly, in these empirical findings, the groups identified to use reinforcement are im general populist parties or candidates. 
It is a central argument in the literature (see introduction) that populist parties fuel and use echo chambers and communication bubbles, a finding which we confirm here using a mathematical and statistical neutral model. However, it is intriguing that not all populists groups leave detectable traces in election data. While Trump in the US, the Brexit, or the German AFD clearly can be identified as candidates/elections/groups where reinforcement played a role, our model does not show a signal for Le Pen in France. 
Marie Le Pen is one of the central persons in contemporary France politics, and classically defined as a far right wing populist. Our method detects derivations of the vote share distributions from the beta distribution, predicted by the zealot model. The results of Le Pen does not show that spatial heterogeneity which is detectable by our method. We offer several non-mutually exclusive explanations. The central organization of French politics and French administrations (for exemple relatively to Germany) may promote homogeneity, thus making the populist signature not observable by our method. The reinforcement for populist parties may take place in social groups that are not spatially segregated. In the later case, the segregation triggered by reinforcement cannot be detected using spatial data. Finally, the social and political structure in France may generate an overall polarization effect which decreases the observable signatures of echo chambers in the votes. In other words, polarization (reinforcement) for ar against a candidate appear to fairly common for several candidates per election as suggested by the analysis of several presidential election results.  
\par\medskip 


The third central finding is the prediction of phase transitions in electoral systems by populist parties and candidates; this prediction is confirmed in data from The Netherlands. The prototypical model in statistical physics exhibiting phase transitions is the Potts model respectively the Glauber  dynamics~\cite{Krapivsky2017,Glauber1963}. Also in application close to election data (twitter data ahead of elections) indicated that this system is close to a phase transition~\cite{Nicolao2019}. As the basis of the Glauber dynamics (so-called low temperature limit) is a majority rule, while the model presented here is based on interactions of single individuals, we think that our model is more appropriate for social dynamics. \\
The social mechanisms that create the phase transitions are clearly visible in the Dutch Catholic People's Party in the 1970s. The data show that this party did divide the population. The religious segregation at that time corresponded to a certain spatial segregation: Catholic and Protestant population tended to separate. Spatial or social segregation is for sure one of the major driver for reinforcement, as this segregation creates a homogeneous environment that minimizes the contact with different opinions. The spatial model shows that also the reverse direction is possible: alignment of the opinion with close neighbors yields locally homogeneous population, such that in consequence different opinions prevail in different regions, with sharp spatial transitions. \par\medskip

\par\bigskip

We interpret our findings in the context of perceptual psychology. The main point of the present study is the question how individuals form their opinion. In that, the various aspects of the information bias plays a role. A focal individual forms his/her opinion based on new, external information, and based upon his/her pre-knowledge resp.\ prejudices. A main aspect of external information is ``framing''. Particularly complex information are often simplified and presented in a (group-specific) representation/interpretation~\cite{Chong2007}. In our model, the effect of framing can be identified with the effect of zealots. A person, however, filters incoming information based on his/her experiences, including his/her social (network) contacts. As a result, we find the ``confirmation bias''. Individuals seem to rather select external information that agrees with his/her own opinion~\cite{Nickerson1998}.  The present study is a hint that the confirmation bias is active, at least on a group level. Our results might indicate that confirmation bias does not only depend on the information {\it per se}, or the mindset of a focal individual, but that also the context of the social environment influences the  readiness to accept new arguments. While the term ``confirmation bias'' mainly focus on an individual and the information at hand, it is likely that information is also rejected as it does not align with the opinion in the close social environment. This interpretation of the confirmation bias/reinforcement resembles the emphasis of ``tradition'' for the development of knowledge in the theory of Feyerabend~\cite{Feyerabend1982}.\\
The effect of confirmation bias/reinforcement, understood in the sense discussed above, is rather strong. It is even possible that the prevailing opinion cannot be predicted based on the parameters - we are faced with a bistable setting. What are the practical implications and consequences of these findings? At the present time, the  society is faced with major tasks as the implication of the climate change~\cite{Newell2014} of the effect of strong inequalities in the society~\cite{Piketty2017}, to name but a few. Even if there is an almost perfect consensus about a topic at the rational level, as it is in the case of the climate change, only the society is the one who can take actions. At that point, e.g.\ climate-change deniers as well as climate-scientists will use framing. The topic is too complex to be discussed in each detail, and not all details may be known. In any case, climate-change deniers tend also to organize in populist groups and parties (Trump, AFD), which clearly establish structures based on the confirmation bias/reinforcement. We have seen that the reinforcement mechanism is strong as individuals that join that group tend to stay within that group. If the climate-change deniers use reinforcement, while the climes-scientists only argument rationally, the rational information pressure has to be high to shift the overall opinion in the society towards the appropriate actions. A recent example, where this mechanisms may have been decisive, is the Brexit: The brexitiers results show a strong reinforcement component, while the model detected rather little reinforcement on the remainers' side. And indeed, the brexitiers have been successful with this strategy. If the mechanisms addressed here are well understood, they have the potential to be used and miss-used for a kind of social engineering. 
\par\medskip

We might speculate about the bases of populism and in particular of reinforcement. We emphasize that what follows is not a direct consequence of the model and the model-based data analysis, but rather few speculative interpretations. In Europe populism has gained importance in recent years  \cite{GuardianPopulism, HawkinsPopulism}. We can speculate that this may be a consequence of neo-liberalism, which gained importance since 1980s. As neo-liberalism tends to strongly emphasizes the predominance of individual values versus societal values, this can lead to weaken the agreement on common values, common ideas, and common communication codes. In this situation, sub-groups or candidates can challenge the accepted communication framework, such as the far right-wing populist party AFD in Germany, breaking away from the well established convention on how to address the Holocaust question in Germany.\\
Another explanation is discussed by Van Reybrouck~\cite{Reybrouck2016}. He suggests that modern democracies tend to become technocracies, and due to the complexity of modern societies and economics, experts do influence politics and policies substantially. Moreover, supra-national organizations (IMF, WTO, UN, EU, ...) and supra-national treaties have become important. However, these organizations are rather complex and opaque structures and more or less democratically legitimized. As such, they are the perfect targets fuelling populist movements ideas of an international corrupted elite. \\
Lastly, there may even be evolutionary mechanisms promoting reinforcement. About 150 000 years ago, modern humans formed small groups or tribes that did compete for food and resources. According to Tomassello~\cite{Tomasello2016}, each of these groups established an ``objective'' moral, which was group-specific. This ``objective'' morale established and enforced norms and standards for the group members that may have guaranteed the (optimal) functioning of the group. Therefore, the tendency of humans to align with their social environment, and therewith the origin of reinforcement, may be even seen as the heritage of behavioural selection in early mankind.


\bibliographystyle{abbrv}
\bibliography{socialSciencesLit}

\begin{appendix}

\section{Model analysis}

We perform a deterministic limit of our model, and a weak.effects limit; the deterministic limit allows for a bifurcation analysis. Accordingly, in the weak-effects limit, we find phase transitions. Moreover, the invariant measure of the weak.effects limit is used in the data analysis.\\
Recall that the model reads (details about the meaning of parameters are stated in the main part of the paper)
\begin{eqnarray}
X_t\rightarrow X_t + 1 &\mbox{ at rate } &  \mu (N-X_t)\,\frac{\vartheta_1 (X_t+N_1)}{\vartheta_1(X_t+N_1)+(N-X_t+N_2)},\\
X_t\rightarrow X_t - 1 &\mbox{ at rate } &  \mu X_t\,\frac{\vartheta_2 (N-X_t+N_2)}{(X_t+N_1)+\vartheta_2 (N-X_t+N_2)}.
\end{eqnarray}

Only the latter case we obtain a limiting ODE.  In order to better understand the consequences of the mechanism proposed, we first consider the deterministic limit. 

\subsection{Deterministic limit}

\par\medskip 
\begin{prop} Let $N_i=n_i\, N$. Then, the deterministic limit 
	for $x(t)=X_t/N$ reads
	\begin{eqnarray}\label{reforceODE}
	\dot x &=& 
	-\mu x\,\frac{\vartheta_2 (1-x+n_2)}{(x+n_1)+\vartheta_2 (1-x+n_2)}
	+
	\mu (1-x)\,\frac{\vartheta_1 (x+n_1)}{\vartheta_1(x+n_1)+(1-x+n_2)}.
	\end{eqnarray}
	For $n_1=n_2=n$ and $\vartheta_1=\vartheta_2$, $x=1/2$ always is a stationary point; this stationary point undergoes a pitchfork bifurcation at $\vartheta_1=\vartheta_2=\vartheta_p$, where 
	\begin{eqnarray}
	\vartheta_p & = & 
	\frac{1-2n}{1+2n}.
	\end{eqnarray} 
\end{prop}
{\bf Proof: }
The rates to increase/decrease the state can be written as
$f_+(X_t/N)$ resp.\ $f_-(X_t/N)$, where (recall that $n_i=N_i/N$)
$$f_+(x) = \mu (1-x)\,\frac{\vartheta_1 (x+n_1)}{\vartheta_1(x+n_1)+(1-x+n_2)},\qquad 
f_-(x) = \mu x\,\frac{\vartheta_2 (1-x+n_2)}{(x+n_1)+\vartheta_2 (1-x+n_2)}.$$
Therewith, the Fokker-Planck equation for the large population size (Kramers-Moyal expansion)\index{Kramer Moyal expansion} reads
$$ \partial_t u(x,t) 
= -\partial_x( (f_+(x)-f_-(x))\,u(x,t)) 
+\frac 1 {2N}\partial_x^2 ( (f_+(x)+f_-(x))\,u(x,t)) 
$$
and the ODE due to the drift term in case of $N\rightarrow\infty$ is given by
$$\frac d{dt} x = f_+(x)-f_-(x).$$
This result establishes eqn.~(\ref{reforceODE}). For the following, let $\vartheta_1=\vartheta_2=\vartheta$. If we also choose $n_1=n_2=n$, we have a neutral model, and $x=1/2$ is a stationary point for all $\vartheta\geq 0$, $n>0$. We find the Taylor expansion of the r.h.s.\ at $x=1/2$ (using the computer algebra package  maxima~\cite{maxima})
\begin{eqnarray*}
	&&\mu^{-1}\frac d {dt} x 
	= -
	x\,\frac{\vartheta (1-x+n_2)}{(x+n_1)+\vartheta (1-x+n_2)}
	+
	(1-x)\,\frac{\vartheta (x+n_1)}{\vartheta(x+n_1)+(1-x+n_2)}\\
	&=& - 2\,\vartheta\,
	\frac{(2\,n+1)\,\vartheta+(2\,n-1)}{ (2\,n+1)\,(\vartheta+1)^2 }\,\,\bigg(x-\frac 1 2\bigg) 
	+
	\frac{32\,\vartheta\,(\vartheta+n-\vartheta^2(n+1))}{(2\,n+1)^3\,(\vartheta+1)^4}\,\, \bigg(x-\frac 1 2\bigg)^3+{\cal O}((x-1/2)^4)
\end{eqnarray*}
For $\vartheta\in(0,1)$, $n>0$, the coefficient in front of the third order term always is non-zero, while the coefficient in front of the linear term becomes zero at $\vartheta=\vartheta_p$. Hence, we have a pitchfork bifurcation at that parameter.
\par\qed\par\medskip

\begin{figure}[h!]
	\begin{center}
		(a)\includegraphics[width=6cm]{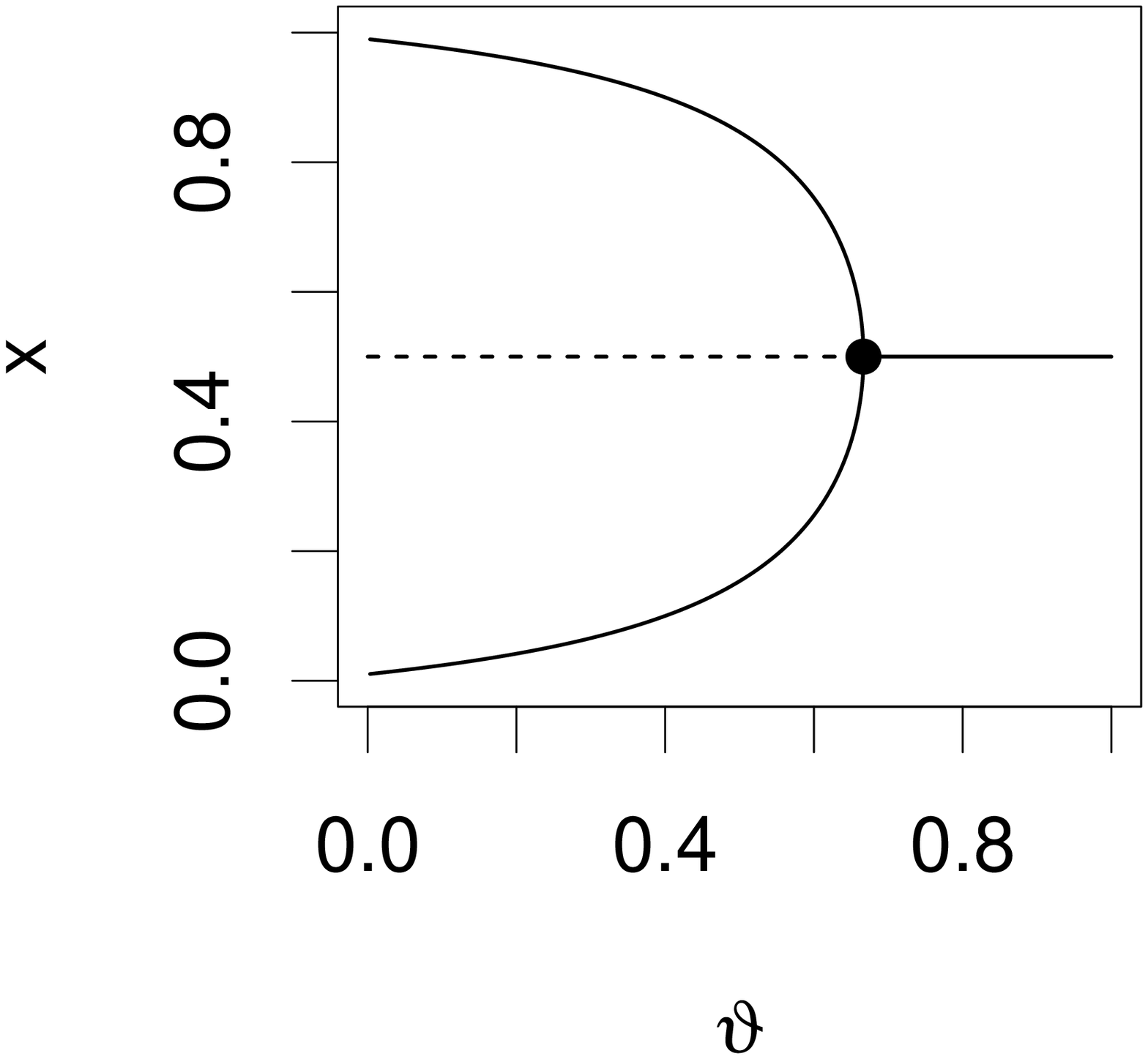}
		(b)\includegraphics[width=6cm]{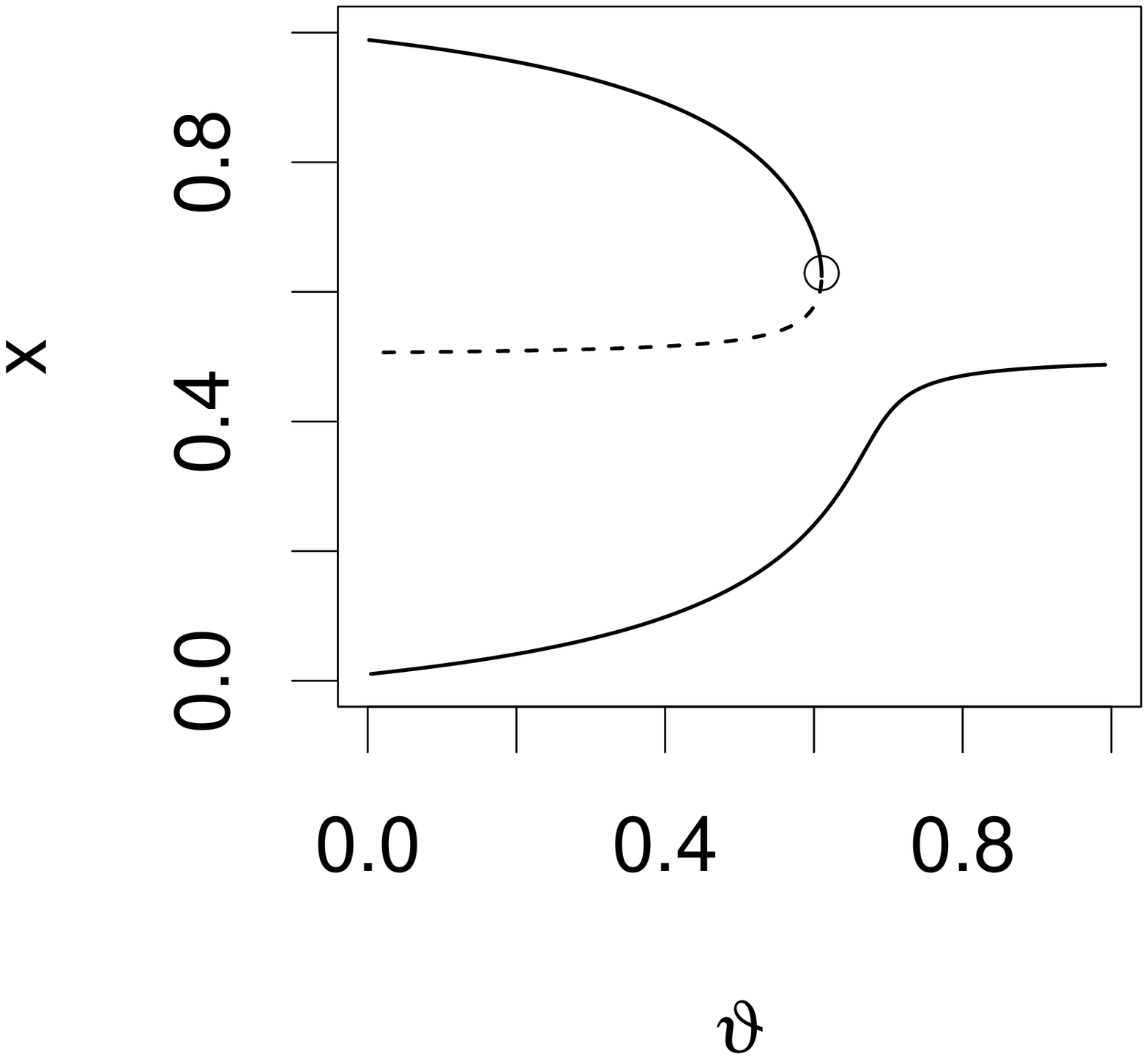}\\
		\vspace*{-1cm}
		(c)\includegraphics[width=6cm]{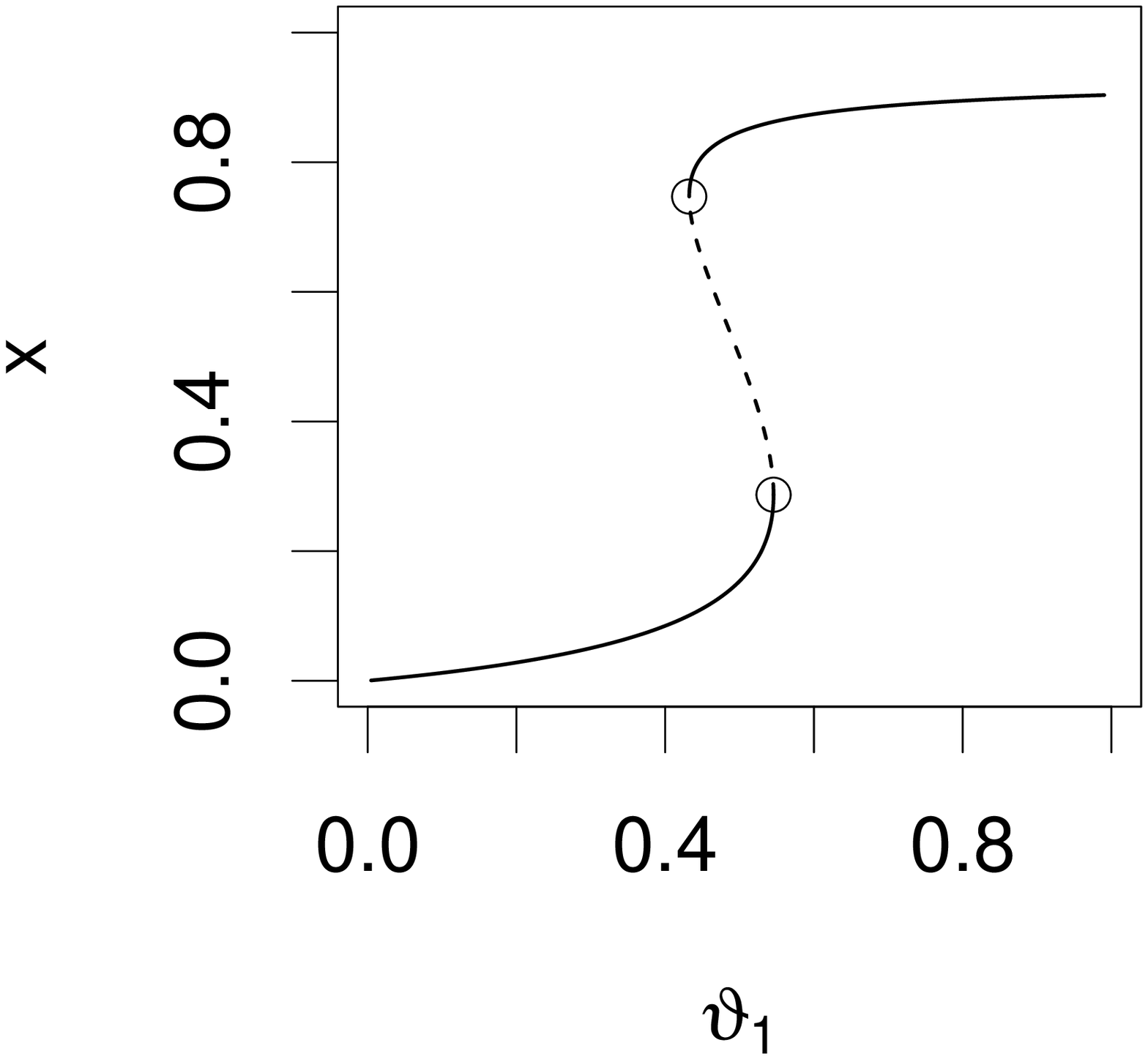}
		(d)\includegraphics[width=6cm]{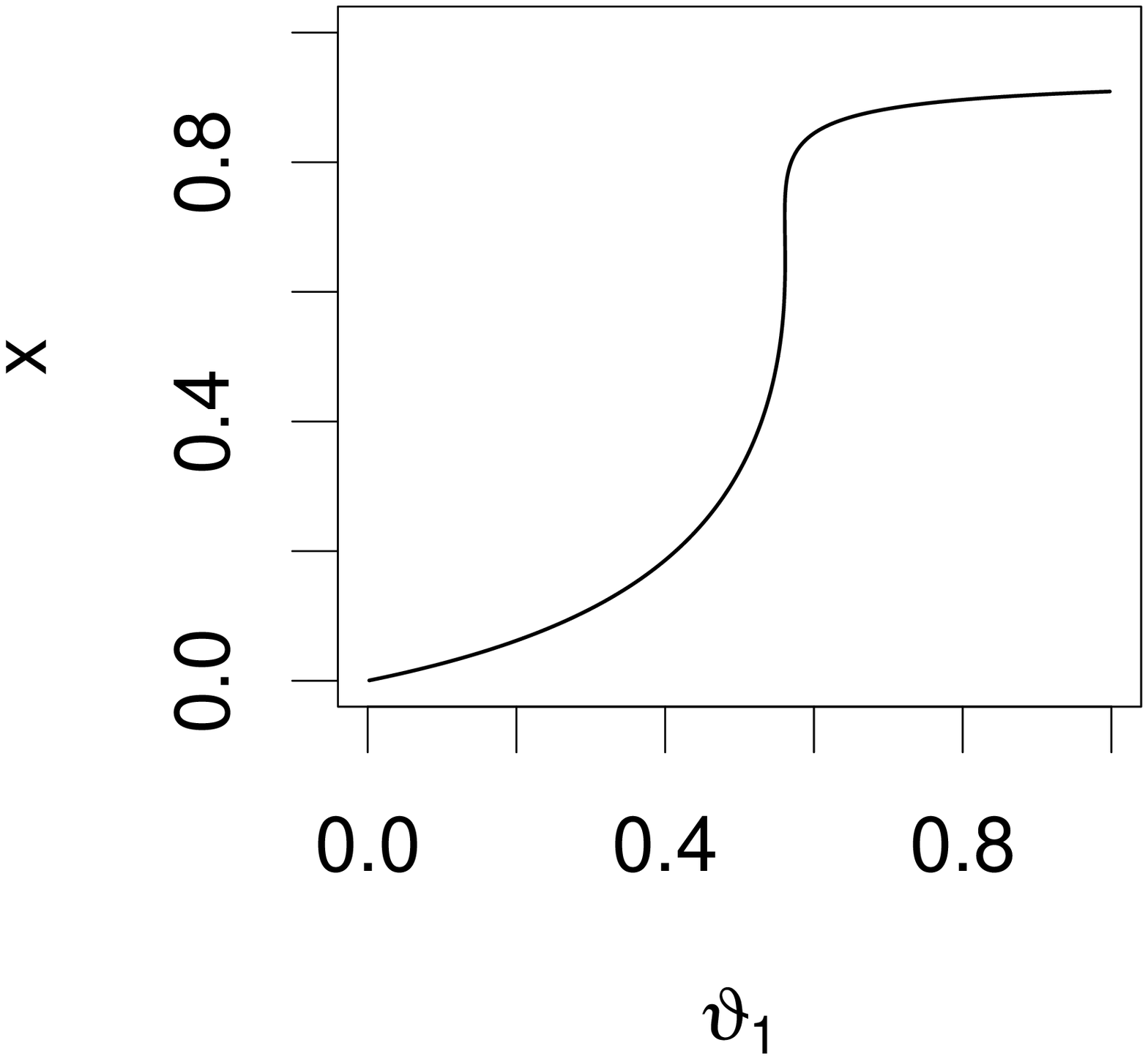}
	\end{center}
	\caption{Stationary points of the reinforcement model over $\vartheta$.  The pitchfork bifurcation in (a) is indicated by a bullet, the saddle-node bifurcations in (b) and (c) are indicated by  open circles. Stable branches of stationary points are represented by solid lines, unstable branches by dotted lines. 
		(a) $n_1=n_2=0.1$, $\vartheta_1=\vartheta_2=\vartheta$, 
		(b) $n_1=0.1$, $n_2=0.105$, $\vartheta_1=\vartheta_2=\vartheta$,  
		(c) $n_1=n_2=0.1$, $\vartheta_2=0.5$, 
		(d) $n_1=0.2$, $n_2=0.02$, $\vartheta_1=1.0$. 
	}
	\label{reinfBifuFig}
\end{figure}

The pitchfork bifurcation is unstable against any perturbation that breaks the symmetry $x\mapsto 1-x$ (Fig.~\ref{reinfBifuFig}). In panel~(a), we have the symmetric case, and find the proper pitchfork bifurcation. Panel~(b) shows the result if the number of zealots only differs slightly, where the reinforcement-parameter for both groups are assumed to be identical. We still find a reminiscent of the pitchfork bifurcation: The stable branches in (b) are close to the stable branches in (a), and also the unstable branches correspond to each other. For the limit $n_2\rightarrow n_1$, panel (b) converges to panel (a). However, the branches are not connected any more but dissolve in two unconnected parts, and the pitchfork bifurcation is replaced by a saddle-node bifurcation. \\
In panel (c) and (d), the upper branch visible in panel~(b) did vanish, and only the lower branch is present. As $\vartheta_2$ is kept constant ($\vartheta_2=0.5$ in panel (c) and $\vartheta_2=0$ in panel (d)) and only $\vartheta_1$ does vary, there is no continuous transition to panel~(a).\par\medskip 

The effect of reinforcement for a given group resembles an increase in the number of the group's zealot. Reinforcement may lead to the dominance of a group. In panel (d), the second group has only $1/10$ of the zealots of the first group, but is able to take over if the members of that group do an extreme reinforcement ($\vartheta_1\ll 1$). However, if the reinforcement of both groups is has a similar intensity and is strong, the mechanism is symmetrical, with a bistable setting as the consequence (panel (a)). 
\par\bigskip

\subsection{Weak effects limit}

We now turn to the second scaling -- the effect of zealots, and also the effect of the echo chambers, are taken to be weak. Under these circumstances, it is possible to find a limiting distribution for the invariant measure of the process.

\begin{theorem}
	Let $N_i$ denote the number of zealots for group $i$, $N$ the population size, and $\vartheta_i=1-\theta_i/N$ the parameter describing reinforcement. In the limit $N\rightarrow\infty$, the density of the invariant measure for the random variable $z_t=X_t/N$ is given by 
	\begin{eqnarray}
	\varphi(x) =  C\,e^{\frac 1 2 (\theta_1+\theta_2)x^2-\theta_1\,x}\,\, x^{N_1-1}\,(1-x)^{N_2-1},\label{reinfDirichDist}
	\end{eqnarray}
	where $C$ is determined by the condition $\int_0^1\varphi(x)\, dx=1$.
\end{theorem}

\begin{figure}[t!]
	\begin{center}
		(a)\includegraphics[width=6cm]{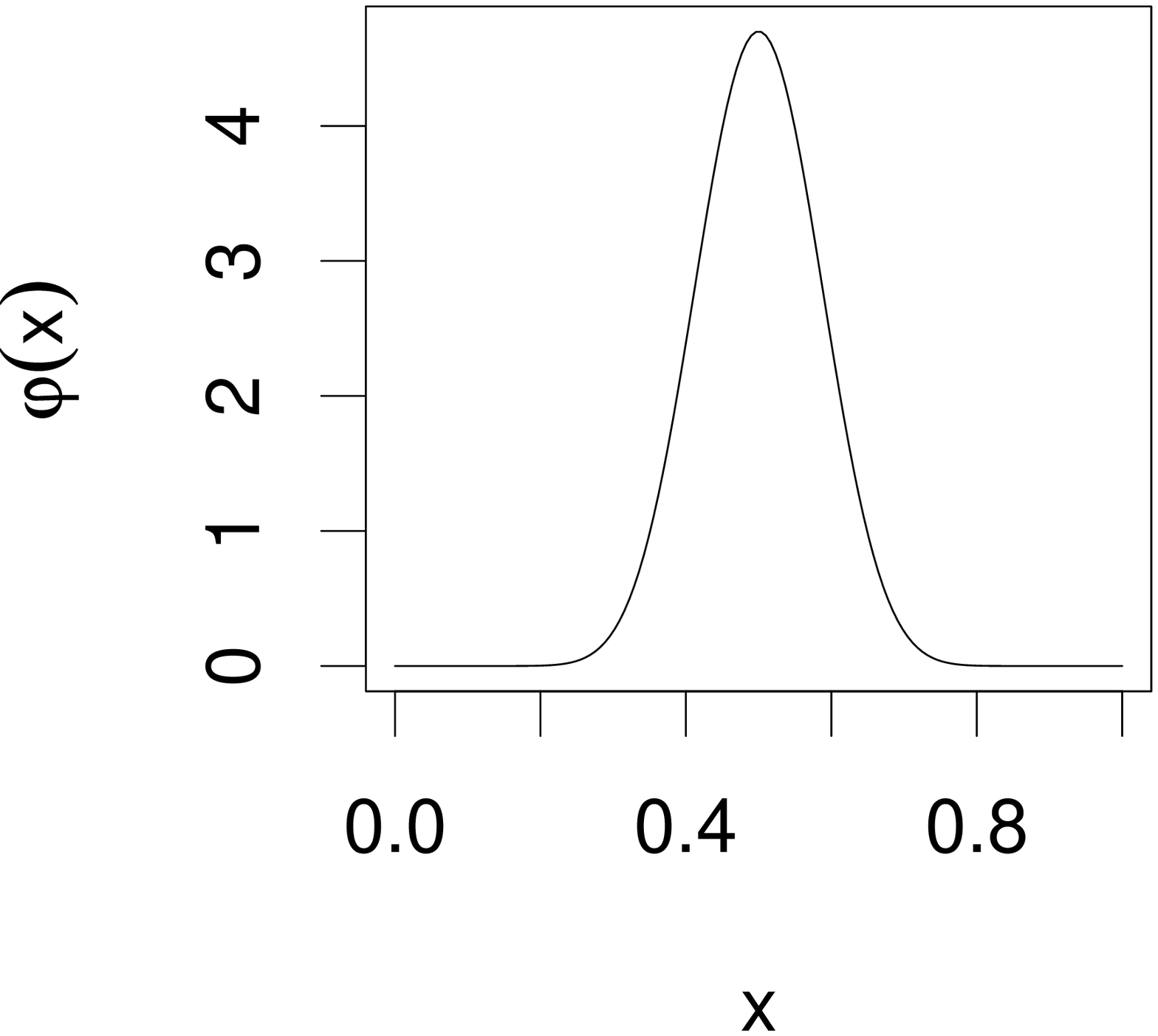}
		(b)\includegraphics[width=6cm]{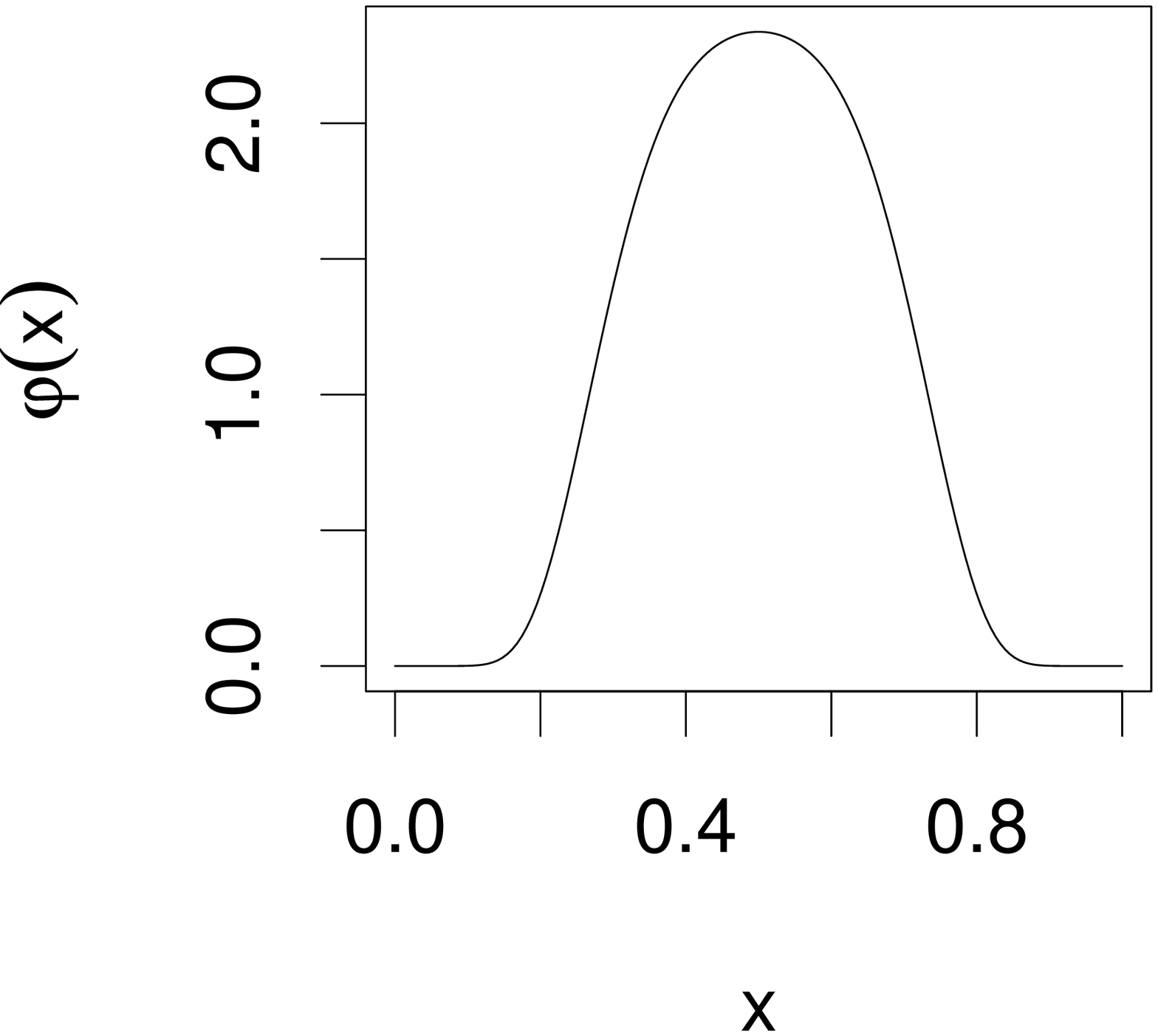}\\
		\vspace*{-1cm}
		(c)\includegraphics[width=6cm]{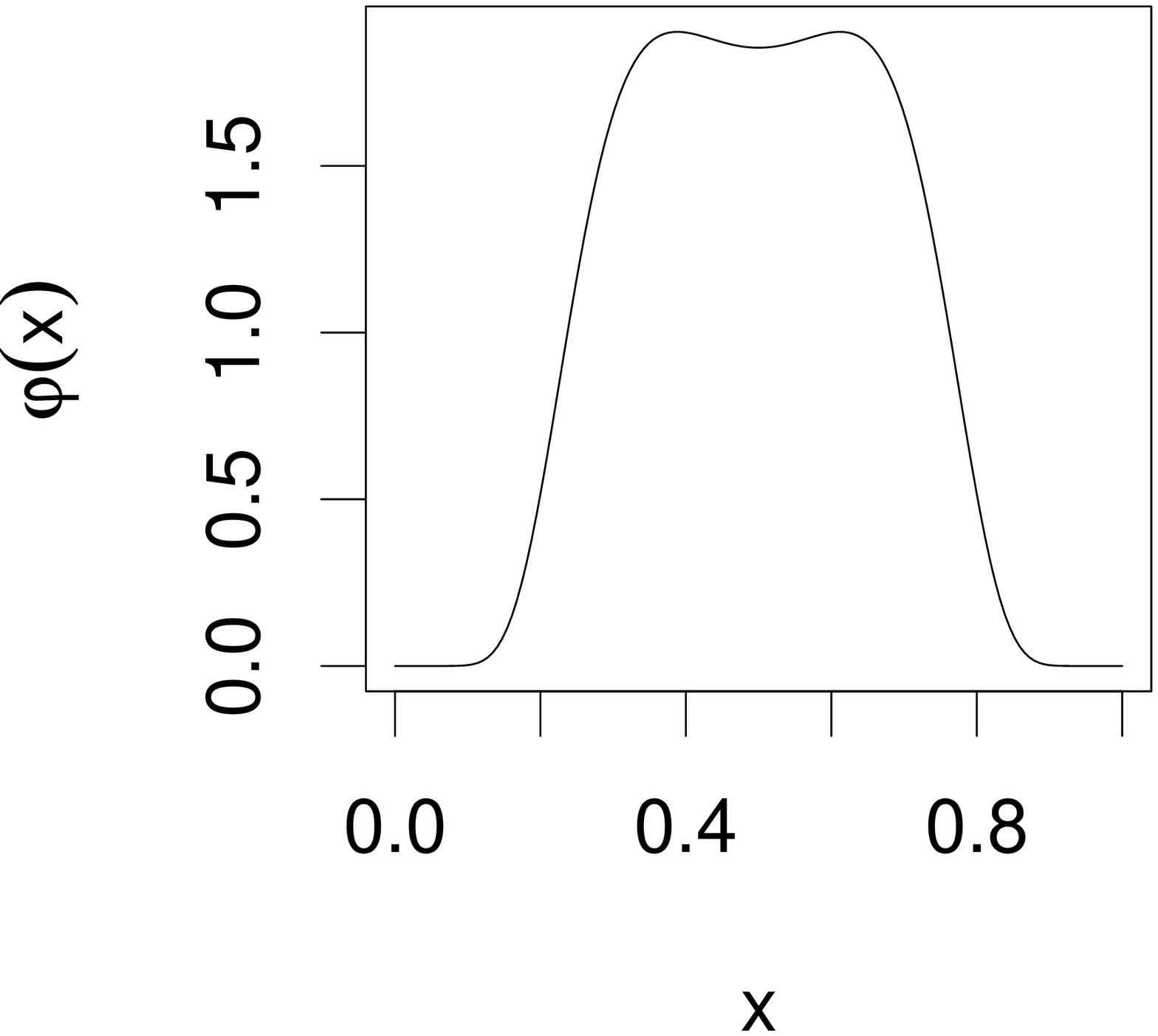}
		(d)\includegraphics[width=6cm]{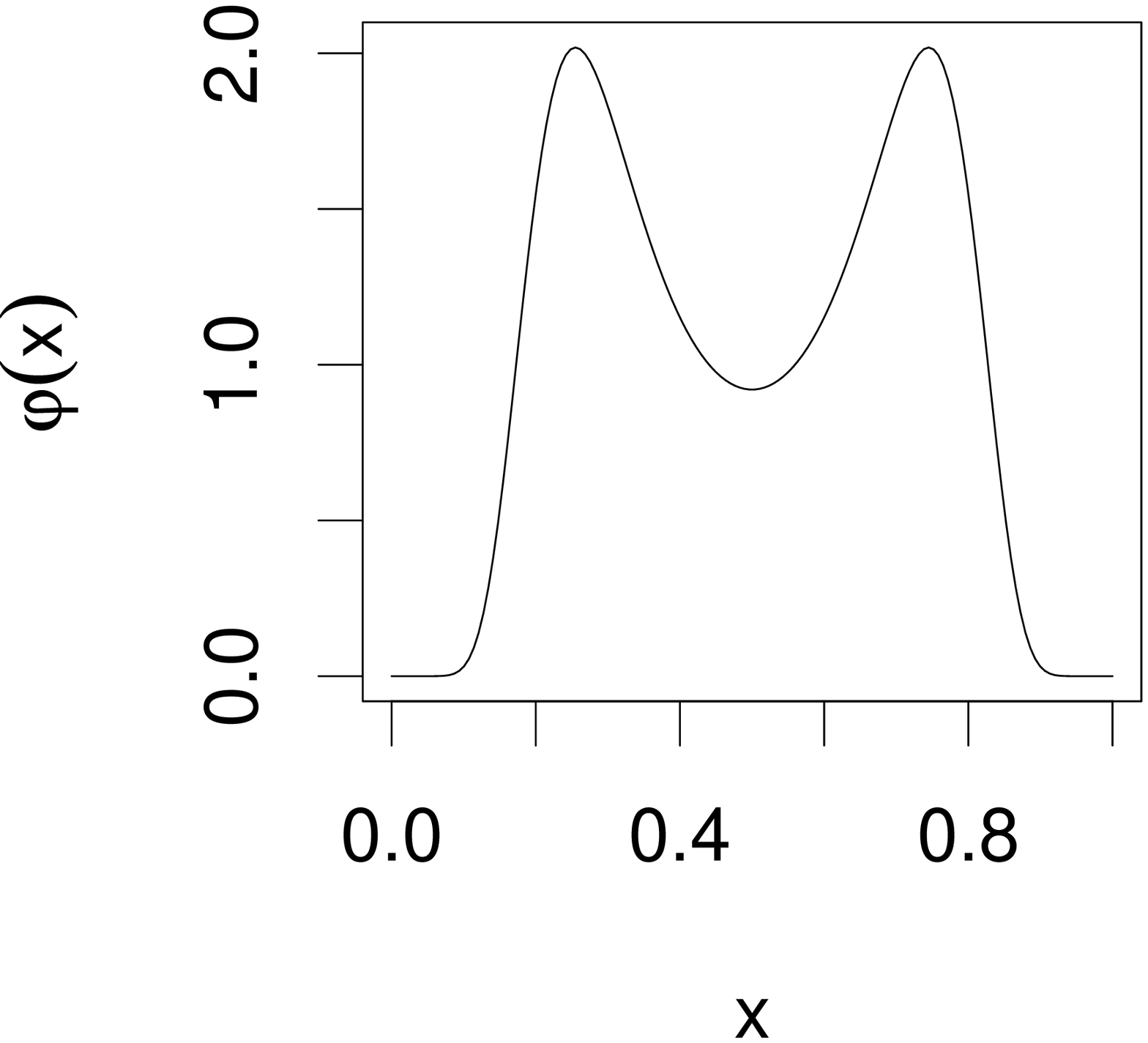}
	\end{center}
	\caption{Invariant distribution, given in eqn.~(\ref{reinfDirichDist}) for $N_1=N_2=20$. We have $\theta_1=\theta_2$, where 
		(a)  $\theta_1=\theta_2=10$, 
		(b)  $\theta_1=\theta_2=70$, 
		(c)  $\theta_1=\theta_2=80$, 
		(d)  $\theta_1=\theta_2=100$.	
	}
	\label{reinfDistribSymmFig}
\end{figure}

\begin{figure}[t!]
	\begin{center}
		(a)\includegraphics[width=6cm]{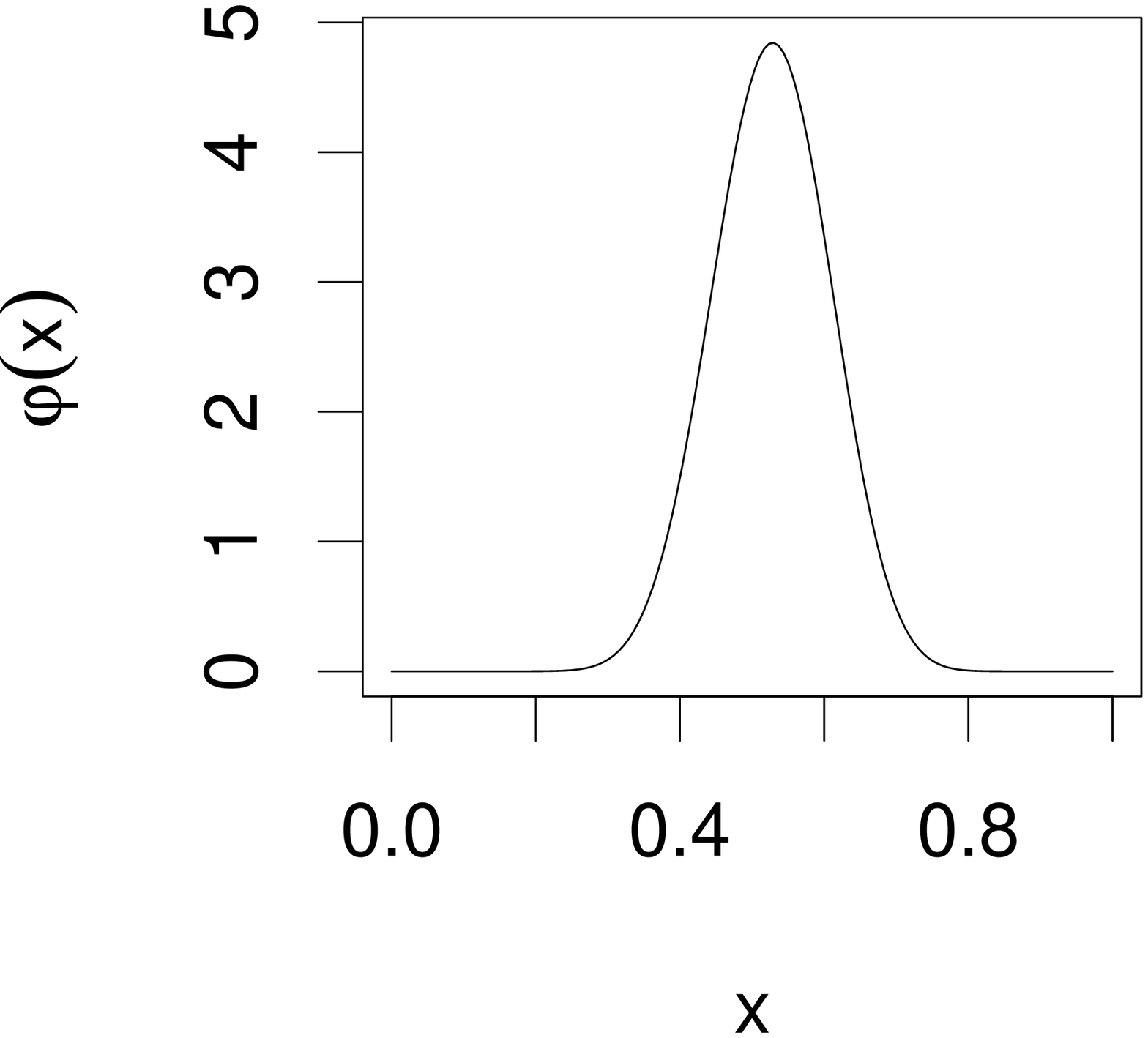}
		(b)\includegraphics[width=6cm]{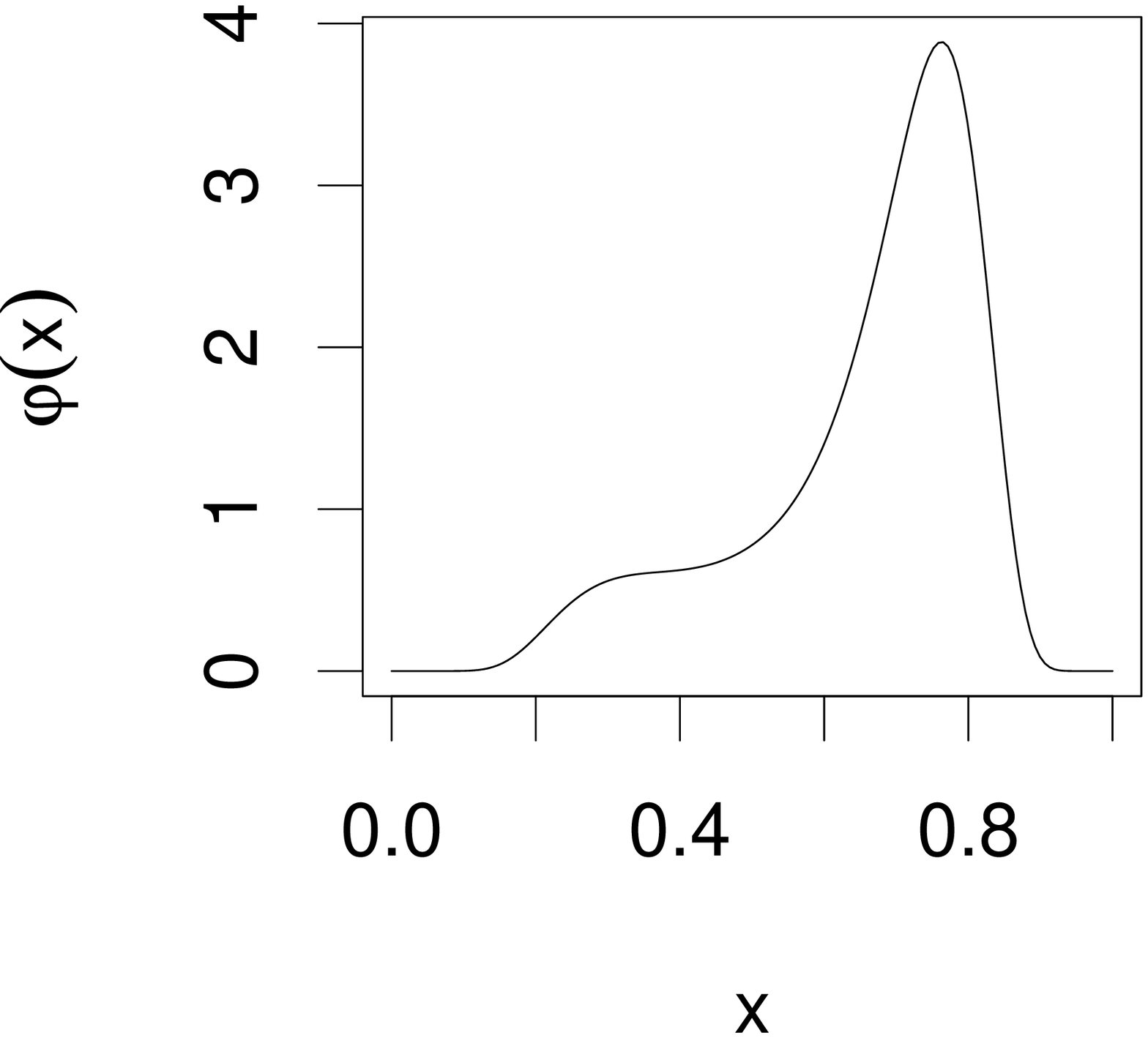}\\
		\vspace*{-1cm}
		(c)\includegraphics[width=6cm]{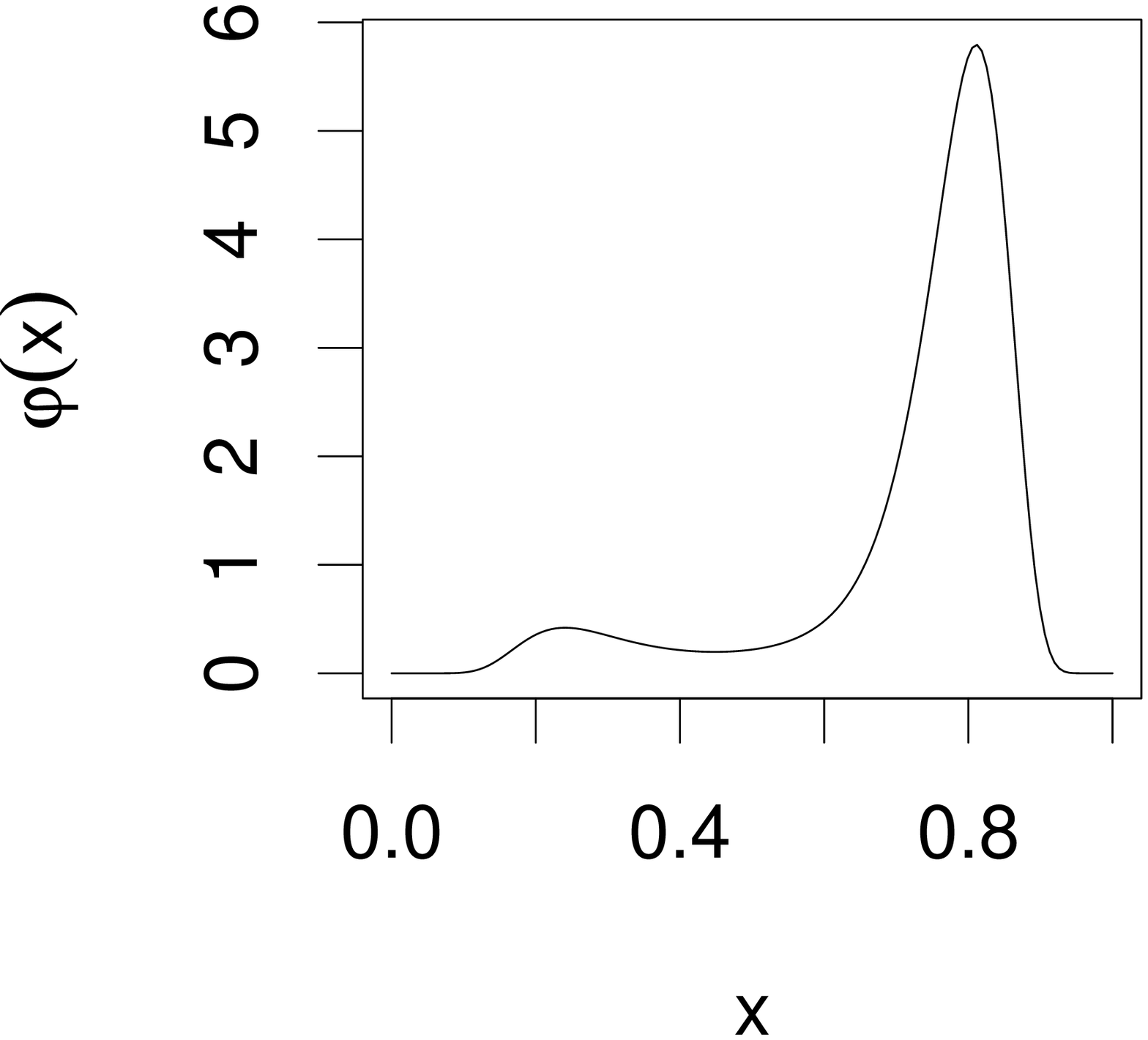}
		(d)\includegraphics[width=6cm]{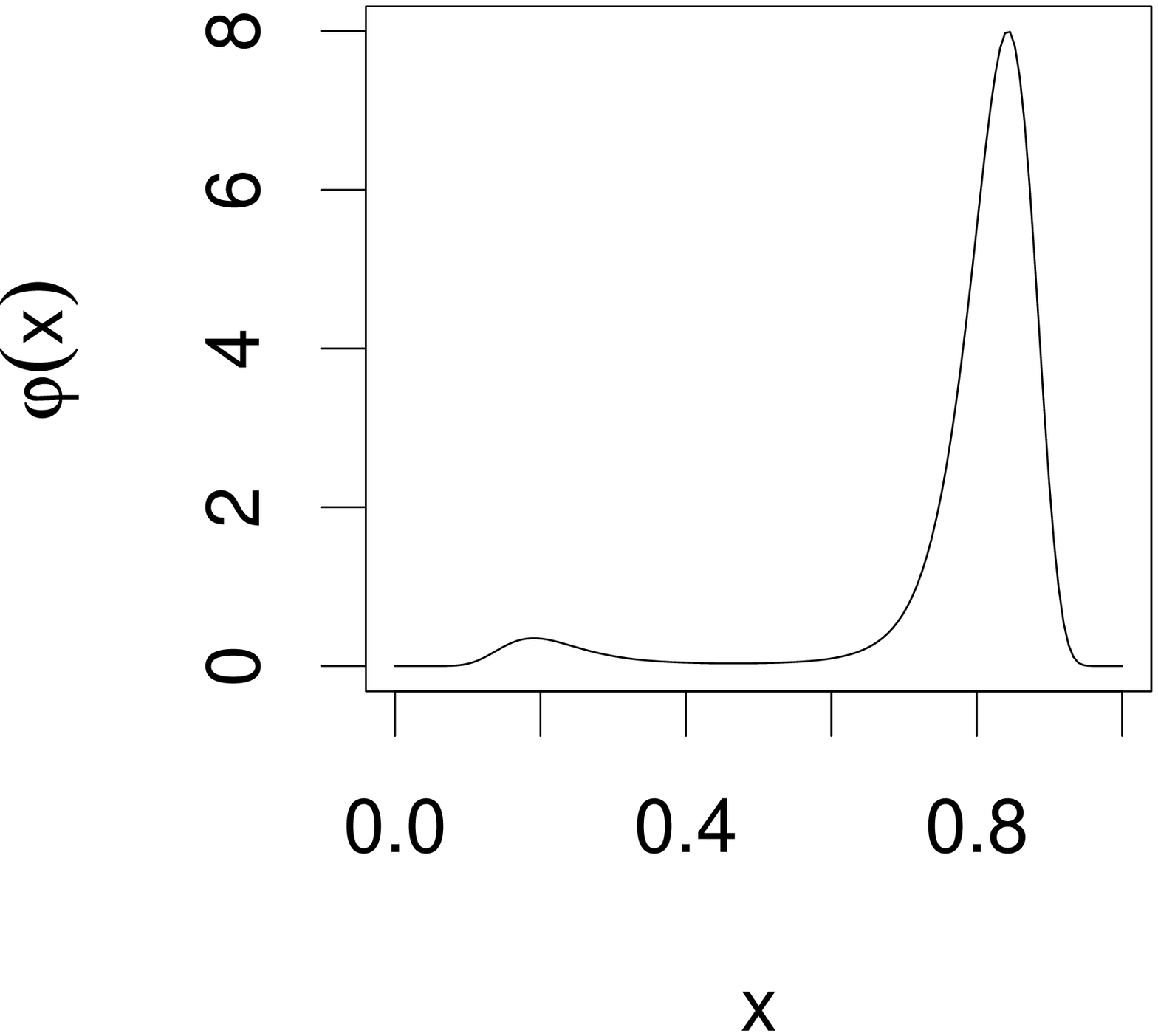}
	\end{center}
	\caption{Invariant distribution, given in eqn.~(\ref{reinfDirichDist}) for $N_1=22$, $N_2=20$. We have $\theta_1=\theta_2$, where 
		(a)  $\theta_1=\theta_2=10$, 
		(b)  $\theta_1=\theta_2=100$, 
		(c)  $\theta_1=\theta_2=120$, 
		(d)  $\theta_1=\theta_2=140$.	
	}
	\label{reinfDistribNonSymmFig}
\end{figure}
{\bf Proof: } We again start off with the Fokker-Planck equation, obtained by the Kramers-Moyal expansion, where we use the scaling $n_i=N_i/N$, and $\vartheta_i$ constant in $N$. Only afterwards, we proceed to the desired scaling.\\
As seen above, the rates to increase/decrease the state can 
be written as
$f_+(X_t/N)$ resp.\ $f_-(X_t/N)$, where (recall that $n_i=N_i/N$)
$$f_+(x) = \mu (1-x)\,\frac{\vartheta_1 (x+n_1)}{\vartheta_1(x+n_1)+(1-x+n_2)},\qquad 
f_-(x) = \mu x\,\frac{\vartheta_2 (1-x+n_2)}{(x+n_1)+\vartheta_2 (1-x+n_2)}.$$
Therewith, the limiting Fokker-Planck equation reads
$$ \partial_t u(x,t) 
= -\partial_x( (f_+(x)-f_-(x))\,u(x,t)) 
+\frac 1 {2N}\partial_x^2 ( (f_+(x)+f_-(x))\,u(x,t)) 
$$
Now we rewrite drift and noise term with the new scaling $n_i=N_i/N$, 
$\vartheta_i=1-\theta_i/N$, where we neglect terms 
of order ${\cal O}(N^{-2})$. We find (using maxima~\cite{maxima}) that ($h:=1/N$)
\begin{eqnarray*}
	&&f_+(x)-f_-(x)\\ 
	&=& 
	\mu (1-x)\,\frac{(1-h\,\theta_1) (x+h\,N_1)}{(1-h\,\theta_1)(x+h\,N_1)+(1-x+h\,N_2)}
	-
	\mu x\,\frac{(1-h\,\theta_2) (1-x+h\,N_2)}{(x+h\,N_1)+(1-h\,\theta_2) (1-x+h\,N_2)} \\
	&=& \mu\bigg(
	[(\theta_1+\theta_2)x-\theta_1]\,x\,(1-x)- (N_1+N_2)\,x+N_1
	\bigg)
	\, h+{\cal O}(h^2),
\end{eqnarray*}
while $h(f_+(x)+f_-(x)) = h\,2\,\mu x(1-x)\, + {\cal O}(h^2)$. If we rescale time, 
$T=\mu\, h\,t$, the Fokker-Planck equation becomes
$$ \partial_T u(x,T) = 
-\,\partial_x\bigg\{\,\,\bigg(
[(\theta_1+\theta_2)x-\theta_1]\,x\,(1-x)- (N_1+N_2)\,x+N_1
\bigg)\,\,u(x,T)\,\,\bigg\}
+ \partial_x^2 \bigg\{x\,(1-x)\, u(x,T)\bigg\}.
$$
For the invariant distribution $\varphi(x)$,  the flux of that rescaled Fokker-Planck equation is zero, that is, 
$$-\bigg(
[(\theta_1+\theta_2)x-\theta_1]\,x\,(1-x)- (N_1+N_2)\,x+N_1
\bigg) \varphi(x) + \frac d {dx} \bigg(x(1-x)\,\varphi(x)\bigg) = 0.$$
With $v(x) =x(1-x)\varphi(x)$, we have 
$$
v'(x) = 
\bigg(
[(\theta_1+\theta_2)x-\theta_1]\,+ \frac {N_1}x-\frac{N_2}{1-x}\,\bigg) v(x)
$$
and hence
$$
v(x) = C\,e^{\frac 1 2 (\theta_1+\theta_2)x^2-\theta_1\,x}\,\, x^{N_1}\,(1-x)^{N_2}
$$
resp.\ 
$$
\varphi(x) = C\,e^{\frac 1 2 (\theta_1+\theta_2)x^2-\theta_1\,x}\,\, x^{N_1-1}\,(1-x)^{N_2-1}
$$
\par\qed\par\medskip

For $\theta_1=\theta_2=0$, we obtain the beta distribution, as we fall back to the zealot model without reinforcement. In the given scaling, the reinforcement is expressed by the exponential multiplicative factor. As $\vartheta_i=1-h\,\theta_i$, and $h=1/N$ is small, one could be tempted to assume that we are in the subcritical parameter range of the reinforcement model only, s.t.\ the distribution does not show a phase transition. As we see next, this idea is wrong.\par\medskip 
Let us first consider the symmetric case, $N_1=N_2=\ul N$, and $\theta_1=\theta_2=\ul \theta$ (see Fig.~\ref{reinfDistribSymmFig}). In that case, the distribution is given by 
\begin{eqnarray*}
	\varphi(x) =  C\,e^{-\ul\theta\,x\,(1-x)}\,\, x^{\ul N-1}\,(1-x)^{\ul N-1}.
\end{eqnarray*}
The function always is symmetric w.r.t.\ $x=1/2$. 
If $\ul \theta$ is small, and $\ul N>0$, we find an unimodal function, with a maximum at $1/2$.
If, however, $\hat\theta$ is increased, eventually a bimodal distribution appears -- we find back the pitchfork bifurcation that we already known from the deterministic limit of the model (Fig.~\ref{reinfBifuFig}, panel~a). 
\par\medskip 
As soon as $N_1\not=N_2$, the symmetry is broken (Fig.~\ref{reinfDistribNonSymmFig}), and we have an 
a situation resembling Fig.~\ref{reinfBifuFig}, panel~(b). In the stochastic setting, however, we have more information: the second branch concentrates only little probability mass, and will play in practice only a minor role (if any at all). Only if $N_1$ and $N_2$ are in a similar range, or the dissimilarity is balanced by appropriate reinforcement parameters, this second branch is able to concentrate sufficient probability mass to gain visibility in empirical data.
\par\medskip 

{\bf Comparison of the reinforcement model and the zealot model.} 
We can use the zealot model or we can use the reinforcement model to fit and interpret election data. The zealot model for two parties yields the beta distribution. The density of the reinforcement model basically consist of a product, where one term is identical with the beta distribution, 
$$x^{N_1-1}\,(1-x)^{N_2-1}
$$
while the second term expresses the influence of reinforcement
$$e^{\frac 1 2 (\theta_1+\theta_2)x^2-\theta_1\,x}.
$$
Only if the data have a shape that is different from that of a beta distribution, the reinforcement component leads to a significantly improved fit. This is given, e.g., in case of a bimodal shape of the data (where at least one maximum is in the interior of the interval $(0,1)$), or if the data have heavy tails. Both properties hint to the fact that the election districts are of two different types: one, where the party under consideration is relatively strong, and one where it is relatively weak. This difference, in turn, can be interpreted as the effect of reinforcement: In some election districts voters agree that the given party is preferable, in others they agree that the party is to avoid. The population is not (spatially) homogeneous, but some segregation - most likely caused by social mechanisms - take place. In that, the data analysis of spatially structured election data (results structured by election districts) based on the reinforcement model is able to detect spatial segregation and the consequences thereof. 
\par\medskip

\section{Data analysis}
\label{DatAna}
\subsection{Methods}
{\it Parameter estimation: } We aim at a  maximum-likelihood estimation  of the parameters of the distribution
\begin{eqnarray*}
	\varphi(x) =  C\,e^{\frac 1 2 (\theta_1+\theta_2)x^2-\theta_1\,x}\,\, x^{N_1-1}\,(1-x)^{N_2-1},
\end{eqnarray*}
where $C$ is determined by  $\int_0^1\varphi(x)\, dx=1$.
The maximum likelihood parameter estimation is somewhat subtle as the distribution 
incorporates exponential terms - in particular, if the parameters become large, the integral $\int_0^1 e^{\frac 1 2 (\theta_1+\theta_2)x^2-\theta_1\,x}\,\, x^{N_1-1}\,(1-x)^{N_2-1}\, dx$ becomes numerically unstable. Therefore, we 
re-parameterize the distribution, defining $\hat\nu$, $\hat s$, $\hat\theta$, and $\hat\psi$ by
\begin{eqnarray}
\theta_1 = \hat s\,\hat\theta\,\hat\psi,\quad 
\theta_2 = \hat s\,\hat\theta\,(1-\hat\psi),\quad 
N_1+1 = \hat s\, (1-\hat \theta)\, \hat\nu,\quad 
N_2+1 = \hat s\, (1-\hat \theta)\, (1-\hat\nu),\quad 
\end{eqnarray}
where $\hat\theta,\hat\psi,\hat\nu\in[0,1]$, 
and $\hat s>0$, with the restriction $\hat s\, (1-\hat \theta)\, \hat\nu>1$, 
and  $\hat s\, (1-\hat \theta)\, (1-\hat\nu)>1$. 
Therewith, the distribution becomes 
\begin{eqnarray*}
	\varphi(x) =  \hat C\,\exp\bigg[\,\,\hat s\,\,\,\bigg(
	\hat \theta (x^2/2-\hat \psi \,x)\,\,
	+  (1-\hat\theta)\, \hat\nu \, \ln(x)
	+ (1-\hat\theta)\, (1-\hat\nu)\, \ln(1-x) + A
	\bigg)\,\,\bigg].
\end{eqnarray*}
Here, $A$ is a constant that can be chosen in dependency on the data at hand. In practice, it is used to avoid an exponent that has a very large absolute number. The constant $\hat C$ is, as before, determined by the fact that the integral is one. This form allows for a reasonable maximum likelihood estimation, given appropriate election data. \par\medskip
{\it Numerical issues:} The model assumes continuous data, while the election data are discrete. Therefore, a vote share of 0 or 1 is possible in the empirical data, but the distribution may have poles for those values. We replace all empirical vote shares below $10^{-5}$ by $10^{-5}$, and similarly, all data above $1-10^{-5}$ by  $1-10^{-5}$. In order to determine the normalization constant of $\varphi(x)$, we do not integrate from $0$ to $1$, but only from $0.001$ to $0.999$.  Furthermore, for numerical reasons, we restrict $\hat s$ by an upper limit, that we mostly define as $1800$.
\par\medskip

{\it Test for reinforcement: } The zealot model and the reinforcement model are nested. In that, we can use the likelihood-ratio test to 
check for the significance of the reinforcement component: If ${\cal LL}_0$ is the log-likelihood for the restricted model ($\theta_1=\theta_2=0$, resp.\ $\hat\theta=0$), and ${\cal LL}$ is that for the full  reinforcement model, we have asymptotically, for a large sample size
\index{likelihood ratio test}
$$ 2({\cal LL}-{\cal LL}_0) \sim \chi^2_{2}$$
That is, twice the difference in the log-likelihoods is asymptotically $\chi^2$ distributed, where the degree of freedom is the number of the surplus parameters (here: $\theta_1$ and $\theta_2$, resp.\ $\hat\theta$ and $\hat\psi$, that is, the degree of freedom is $2$). \\
Additionally, we use the Kolmogorov-Smirnov test to find out if the model-distribution of either the full reinforcement mode, or the zealot model ($\theta_1=\theta_2=0$, resp.\ $\hat\theta=0$) agrees with the empirical distribution. If both models are in line with the data, then the reinforcement component will rather not add to the interpretation of the data, if only the reinforcement model approximates the data well (or, at least, much better than the zealot model), we can expect that it is sensible to take the reinforcement component into account.
\par\medskip

\FloatBarrier
\subsection{Details -- US}
In the US,  the candidates for the presidential elections are determined by the ``primary elections''. We do not consider them, but only the presidential elections themselves. 
The election of the president happens indirectly via an Electoral Collage.
Each state nominates a certain number of delegates. In most states, a winner-take-all system is established.  If no candidate receives the majority of the votes, he Congress will elect a candidate.\par\medskip 

The data set used is provided by the Havard Univ., and contains the data on county-level for the elections 2000-2016,\\
\url{https://doi.org/10.7910/DVN/VOQCHQ}, file {\tt countypres\_2000-2016.csv}.
\par\medskip 

\begin{figure}[h!]
	\begin{center}
		\includegraphics[width=4.5cm]{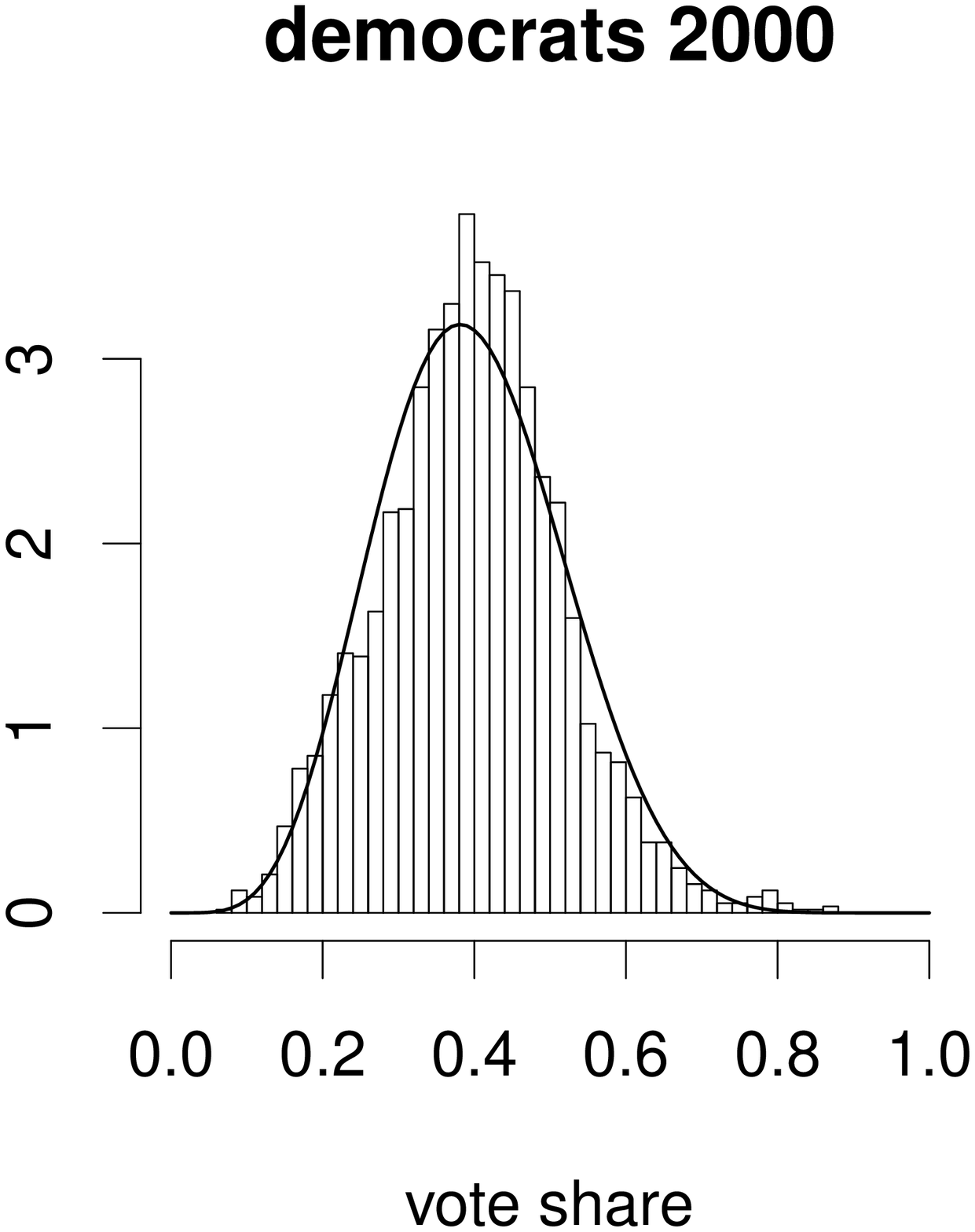}
		\includegraphics[width=4.5cm]{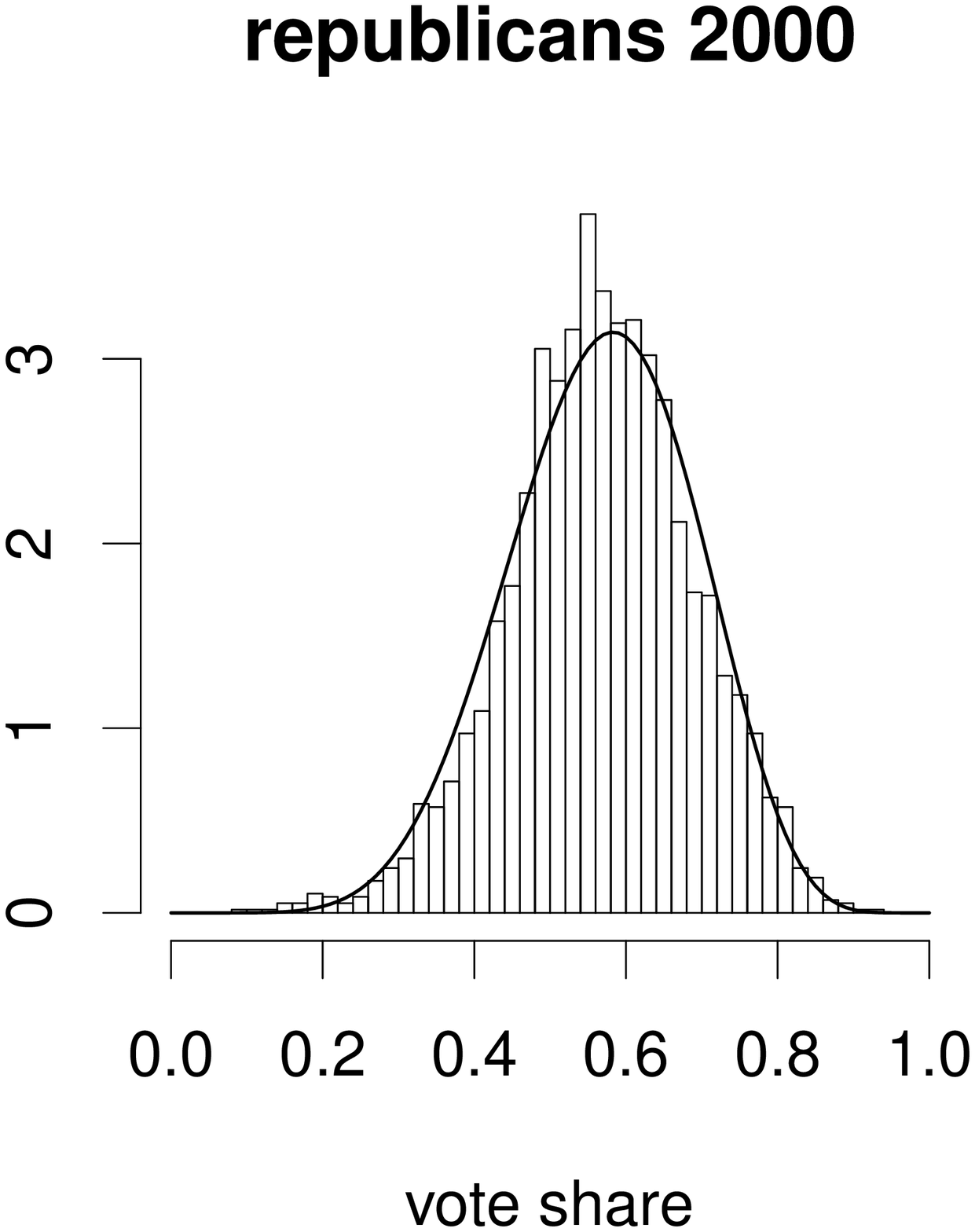}
		\includegraphics[width=4.5cm]{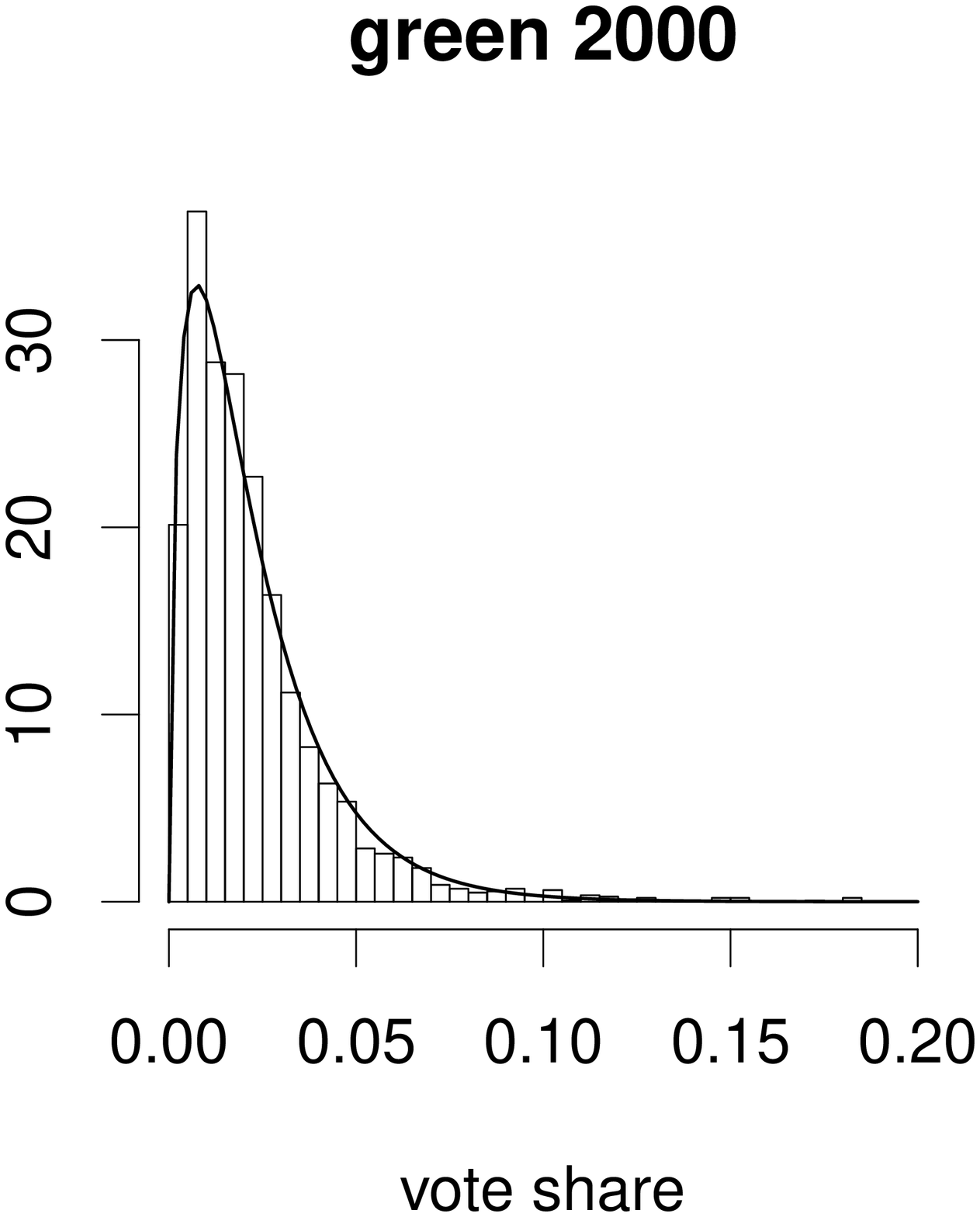}\\
		\includegraphics[width=4.5cm]{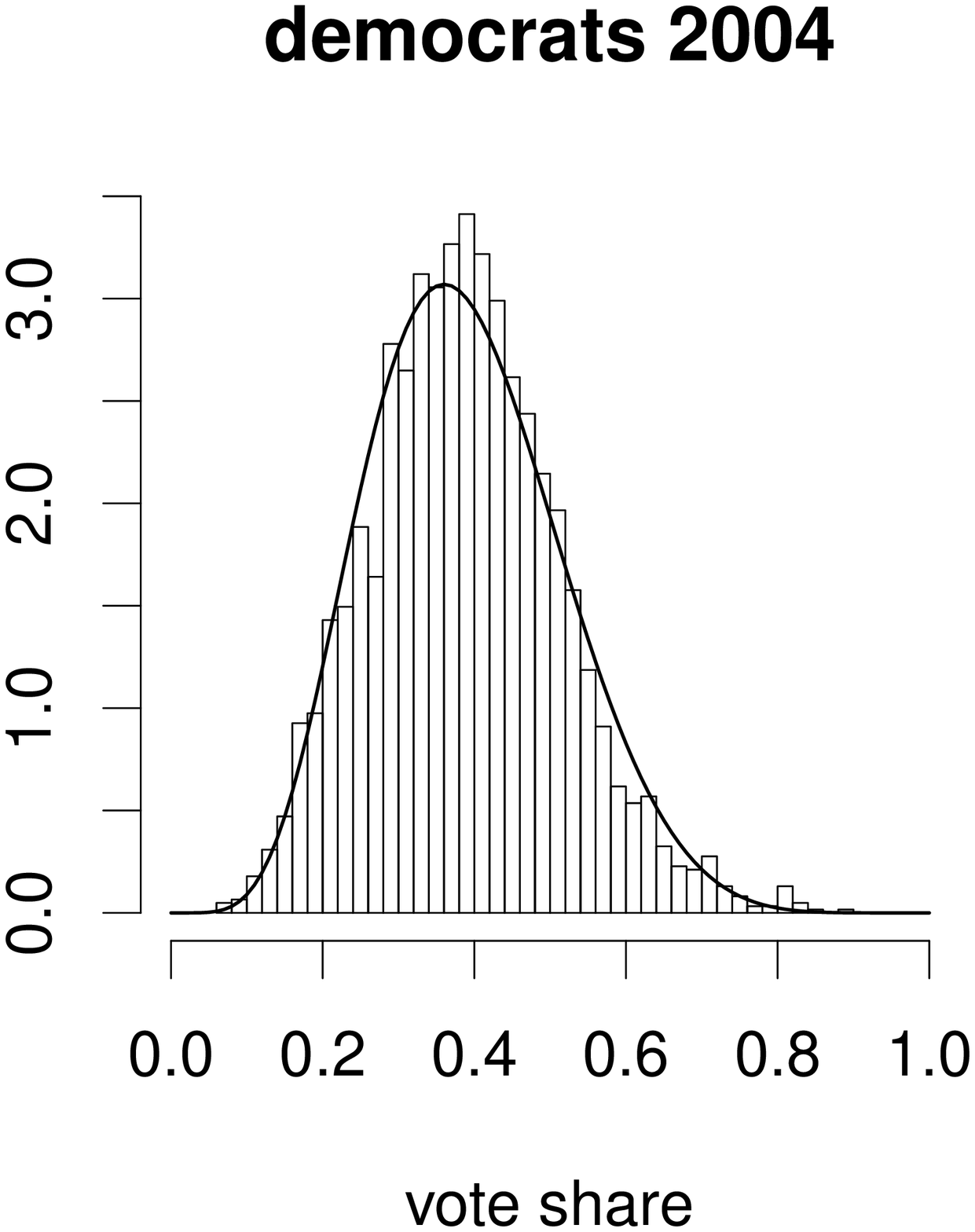}
		\includegraphics[width=4.5cm]{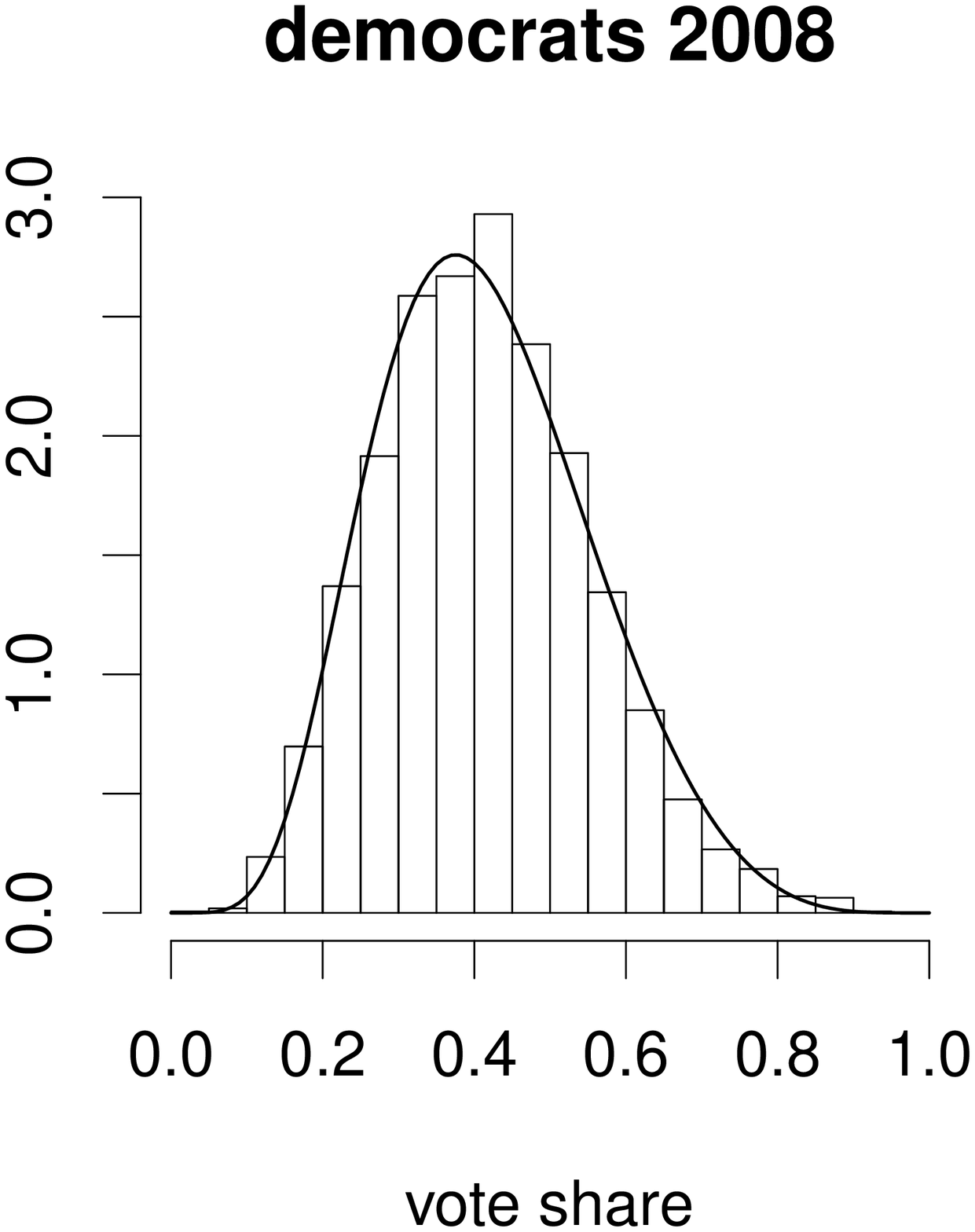}
		\includegraphics[width=4.5cm]{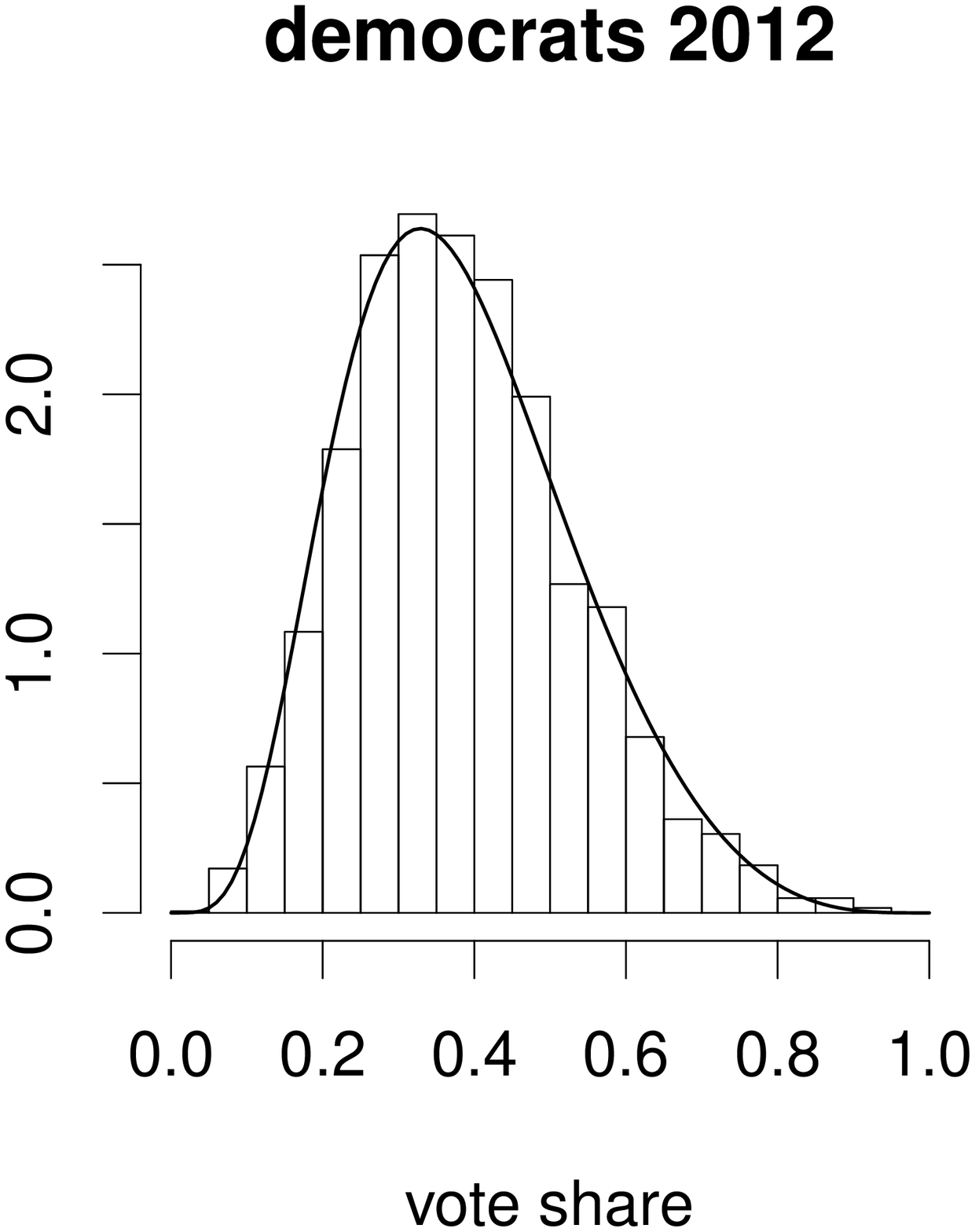}\\
		\includegraphics[width=4.5cm]{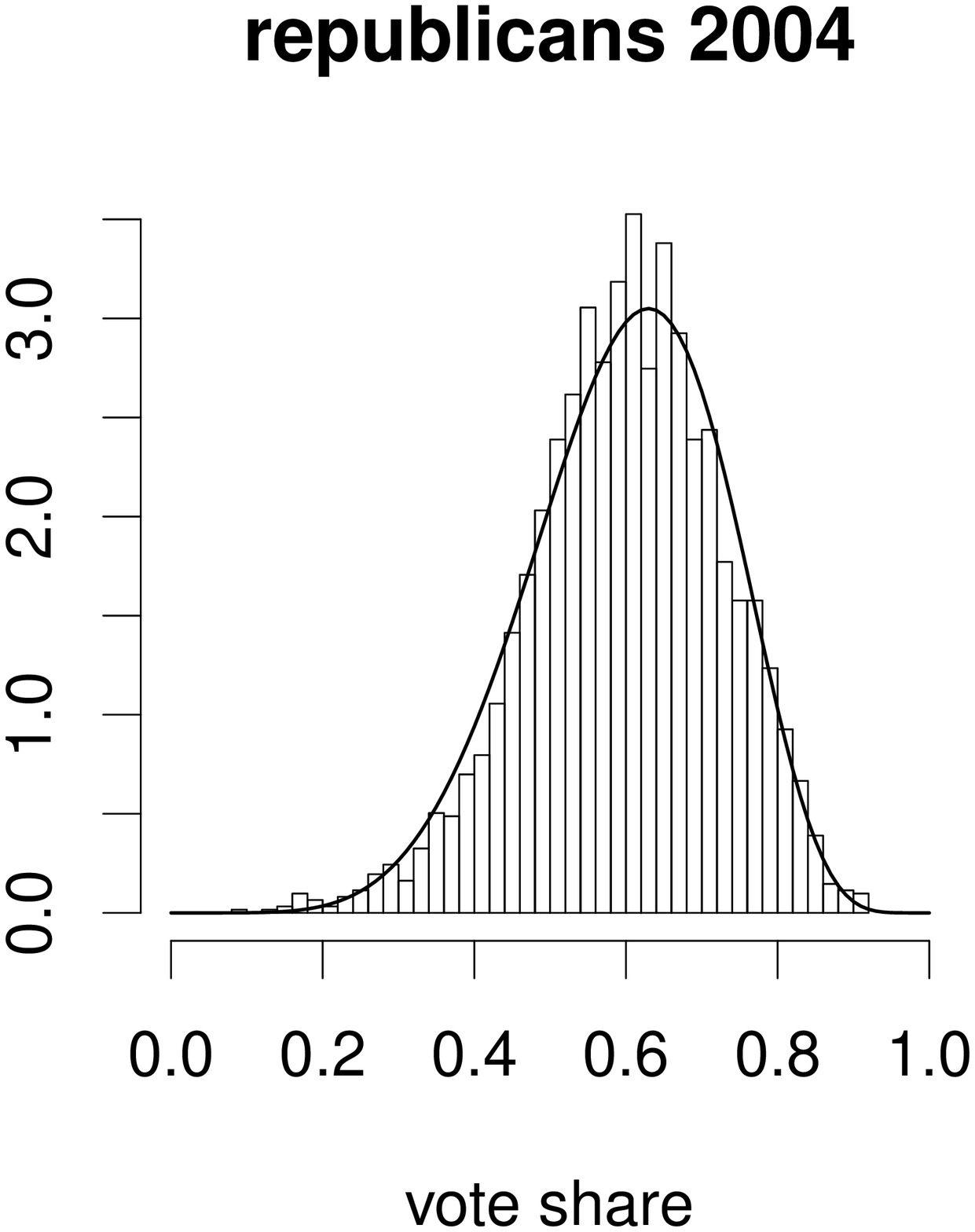}
		\includegraphics[width=4.5cm]{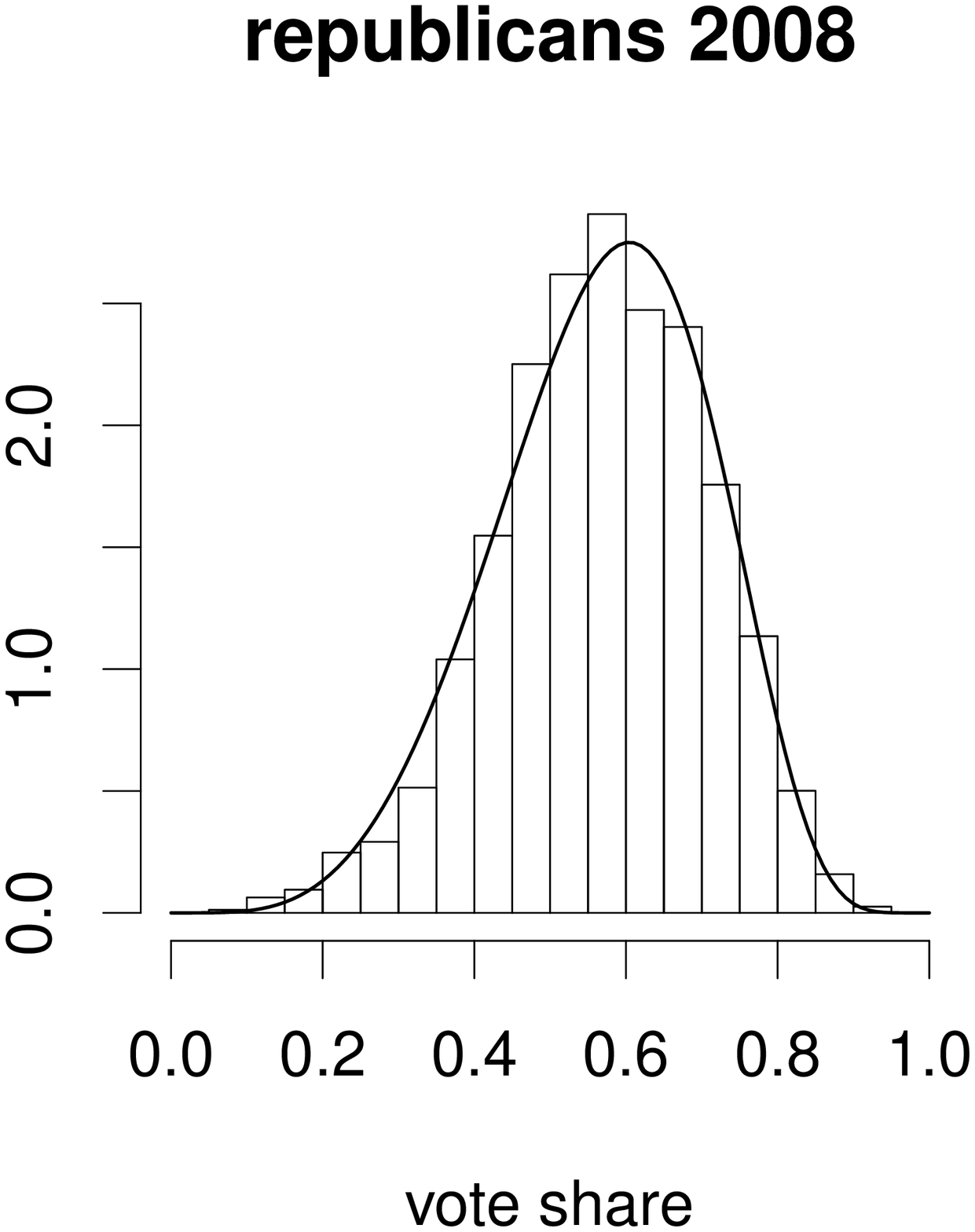}
		\includegraphics[width=4.5cm]{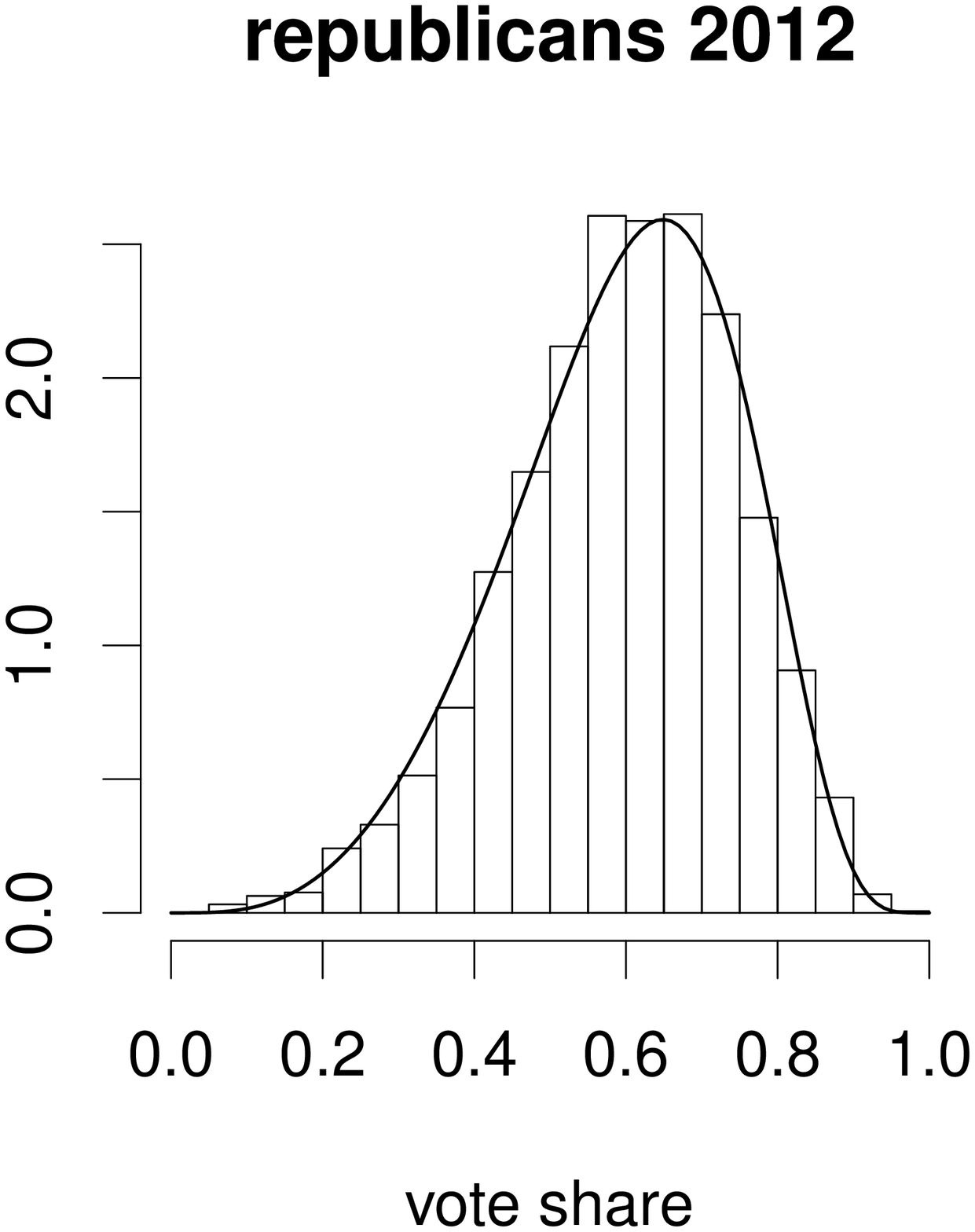}\\
		\includegraphics[width=4.5cm]{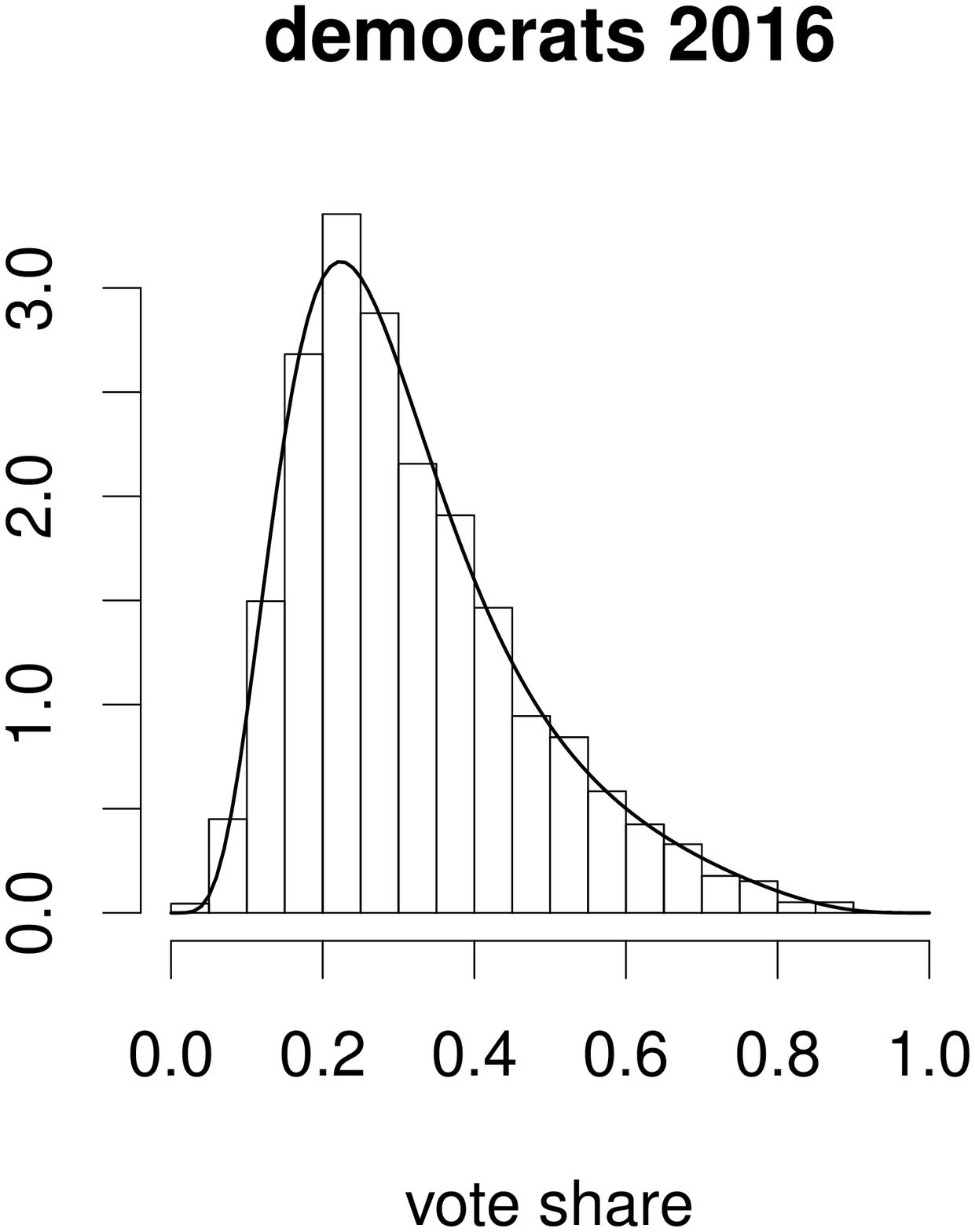}
		\includegraphics[width=4.5cm]{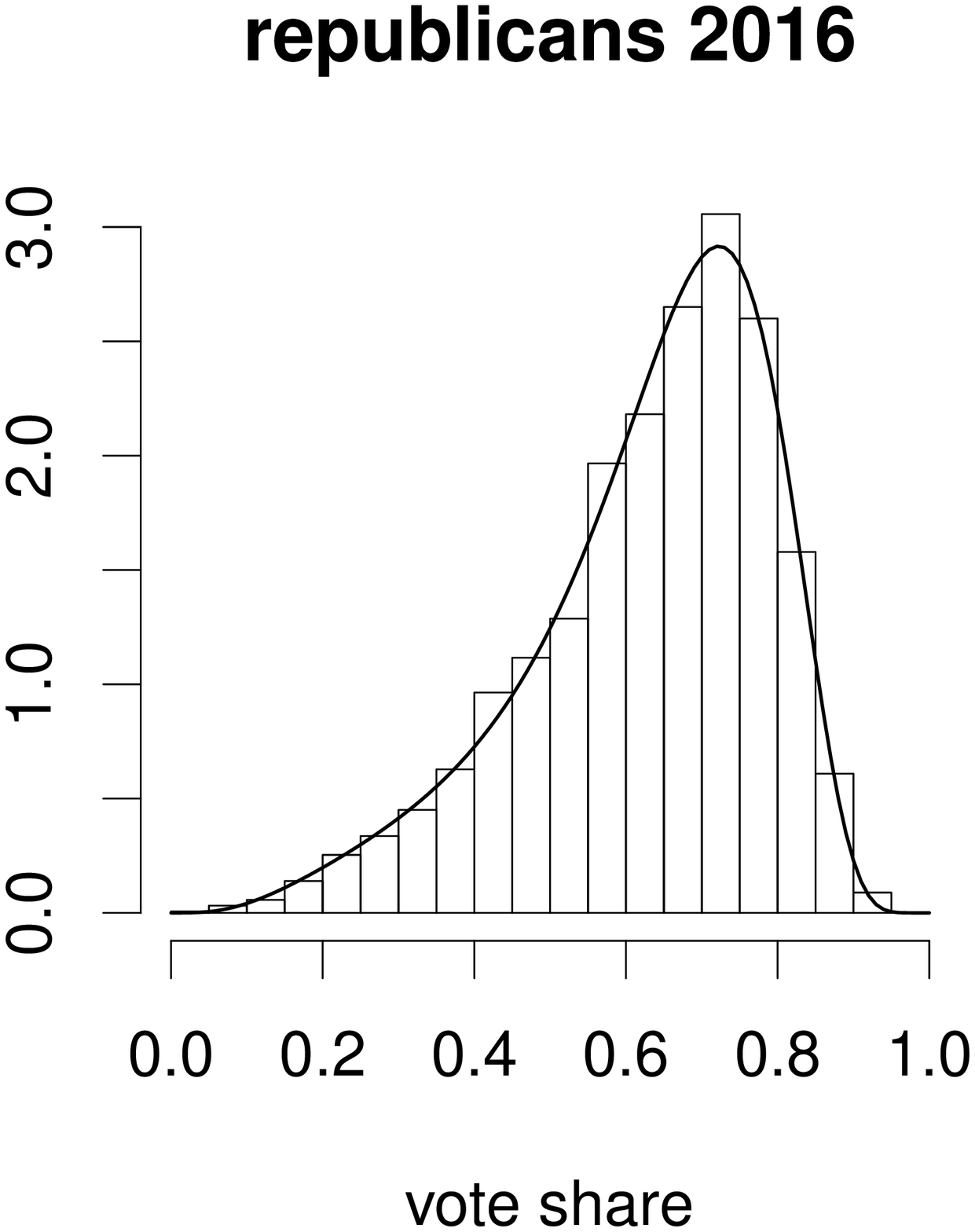}
	\end{center}
	\caption{ Distribution of the vote share of republicans/democrats in 2000-2016 presidential elections. For 2000, we also show the result for the green candidate (note the difference in the scaling of the x-axis here). }
	\label{usPresi20042016}
\end{figure}
\par\medskip

\begin{table}[!h]
	\begin{adjustbox}{width=\columnwidth,center}
		\begin{tabular}{ll|llll|l|lll}
			year & party & $\hat\nu$ & $\hat\theta$ & $\hat\psi$ & $\hat s$ & $\theta_2$ & $p_{ll}$ & $p_{ks}$ (Reinf) &$p_{ks}$ (beta)\\
			\hline
			2000  &  green  & 0.035  &  0.920  & 0.33  &  196.5  & 120.9  & $<$1e-10  &  0.043  &  0.0011 \\
			\hline
			2000  &  republicans  & 0.515&  0.323  & 6.61e-05&  21.6  & 6.97    & 0.86     &  0.02 &  0.02 \\
			2000  &  democrats  & 0.385  &  0.0272 & 0.99993 &  14.0  & 2.51e-05& 1        &  0.007&  0.007 \\
			2004  &  republicans  & 0.485&  0.4971 & 6.61e-05&  26.6  & 13.2    & 0.003    &  0.20 &  0.17 \\
			2004  &  democrats  & 0.474  &  0.435  & 0.99993 &  22.9  & 0.00066 & 0.03     &  0.10 &  0.14 \\
			2008  &  republicans  & 0.399&  0.588  & 6.61e-05&  29.6  & 17.5    & 9.65e-07 &  0.26 &  0.09 \\
			2008  &  democrats  & 0.579  &  0.581  & 0.99996 &  27.4  & 0.00066 & 7.4e-07  &  0.17 &  0.08 \\
			2012  &  republicans  & 0.404&  0.624  & 6.61e-05&  25.3  & 15.8    & 1.5e-10  &  0.43 &  0.009 \\
			2012  &  democrats  & 0.561  &  0.610  & 0.99993 &  23.0  & 0.0009  & 1.6e-09  &  0.39 &  0.012 \\
			2016  &  republicans  & 0.327&  0.793  & 0.207   &  60.0  & 37.7    & $<$1e-10 &  0.88 & $<$1e-10 \\
			2016  &  democrats  & 0.529  &  0.801  & 0.660   &  55.0  & 14.9    & $<$1e-10 &  0.36 &  $<$1e-10 
		\end{tabular}
	\end{adjustbox}
	\caption{Estimated parameters for the two parties in the eight elections. $p_{ll}$ is the result of the likelihood ratio test for the significance of the reinforcement component; $p_{ks}$ is the result of the Kolmogorov-Smirnov-test for the question of the empirical cumulative distribution differs significantly from the cumulative  distribution of the model (either the reinforcement model, or zealot model with the beta distribution). }\label{USreinfParaEst}
\end{table}


\FloatBarrier

\subsection{Details -- Brexit}
In the Brexit referendum (23 June 2016), each voter had the choice ``remain'' or ``leave''. The election was equal, each vote was counted directly. With 51.89\% (and 72.2\% turnout rate), the outcome has been ``leave''. \par\medskip 

The data follow well the reinforcement model (KS, $p= 0.75$), but the zealot model does not fit nicely (KS, $p=0.0009$). The likelihood-ratio-test on the null hypothesis that remainders and brexitiers have the same amount of reinforcement is rejected at $p<$1e-10.\par\medskip

The data set used is provided by the British Government, and contains data on election district level,\\
\url{https://www.electoralcommission.org.uk/who-we-are-and-what-we-do/elections-and-referendums/past-elections-and-referendums/eu-referendum/results-and-turnout-eu-referendum}, file {\tt EU-referendum-result-data.csv}.
\par\medskip 

\begin{figure}[h!]
	\begin{center}
		\includegraphics[width=7cm]{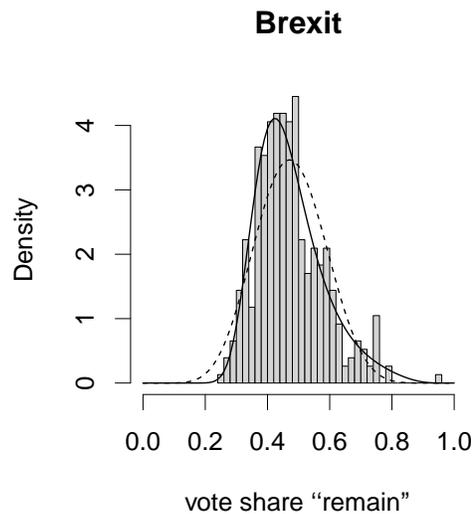}
		\caption{ Distribution of the share of votes for remain in the UK Brexit election with fit of the reinforcement model (solid line) and the zealot model (dashed line).}
	\end{center}
	\label{Figure2}
\end{figure}

\begin{table}[h!]
	\begin{center}
		\begin{tabular}{llll|ll|lll}
			$\hat\nu$ & $\hat\theta$ & $\hat\psi$ & $\hat s$ & $\theta_1$ & $\theta_2$ & $p_{ll}$ & $p_{ks}$ (Reinf.) &  $p_{ks}$ (beta)\\
			\hline
			0.87  &  0.76  & 0.99996  &  254.5  & 193.0  & 0.0084    &  $<$1e-10 & 0.75  &  0.0009  
		\end{tabular}
	\end{center} 
	\caption{Parameter for the Brexit referendum, ``remainers''. For symmetry reasons, the parameters for ``leave'' are identical, but $\hat\psi_{leave}=1-\hat\psi_{remain}$, and therewith $\theta_{2,leave}=\theta_1 = 193$. }
\end{table}
\FloatBarrier

\subsection{Details -- Germany}
Each voter has two votes in the elections for the German parliament: a ``first'' and a ``second'' vote. With the first vote, a candidate in the election district can be selected. The candidate with the most votes will be send into the parliament. With the second vote, a party is selected. The vote share of the parties determines the number of seats in the parliament, where a party has to overcome a 5\% threshold. IN the analysis, only the second votes are used.\par\medskip 
The data are provided by the German Government,\\
\url{https://www.bundeswahlleiter.de/en/bundeswahlleiter.html}, file {\tt btw17\_kerg.csv}.

\begin{table}[h!]
	\begin{center}
		\begin{tabular}{l|llll|l|lll}
			party & $\hat\nu$ & $\hat\theta$ & $\hat\psi$ & $\hat s$ & $\theta_2$ & $p_{ll}$ & $p_{ks}$ (Reinf.) &  $p_{ks}$ (beta)\\
			\hline
			CDU  & 0.35  &  0.34      & 0.67      &  108.5  & 12.4   & 1  &  0.72  &  0.79 \\
			SPD  & 0.18      &  0.30      & 6.61e-05  &  63.5   & 19.0   & 0.99  &  0.19  &  0.20 \\
			AFD  & 0.07  &  0.77      & 6.61e-05  &  792.5  & 609.0  & 8.7e-11  &  0.29  &  0.0009 \\
			FDP  & 0.14   &  0.31      & 1.00      &  147.5  & 0.003  & 1  &  0.55  &  0.55 \\
			die linke  & 0.05&  0.78      & 6.61e-05  &  583.5  & 453.7  & 3.4e-09  &  2.11e-06  &  2.93e-08 \\
			Gruenen  & 0.07  &  0.26      & 6.61e-05  &  69     & 18.2   & 1  &  0.87  &  0.84 \\
			CSU.  & 0.37      &  6.75e-05  & 6.61e-05  &  35.6   & 0.002  & 1  &  0.004  &  0.004 \\
		\end{tabular}
	\end{center} 
	\caption{Results for whole Germany, 2017. }\label{tabReinfGermAll}
\end{table}

\begin{table}[h!]
	\begin{center}
		\begin{tabular}{l|llll|l|lll}
			party & $\hat\nu$ & $\hat\theta$ & $\hat\psi$ & $\hat s$ & $\theta_2$ & $p_{ll}$ & $p_{ks}$ (Reinf.) &  $p_{ks}$ (beta)\\
			\hline
			CDU        & 0.43  &  0.74      & 0.50      &  575.5  & 212.5  & 0.31 &  0.41  &  0.78 \\
			SPD         & 0.21  &  0.09      & 6.61e-05  &  53.5   & 4.87   & 1    &  0.17  &  0.17 \\
			AFD         & 0.13  &  0.29      & 1.00      &  153.5  & 0.003  & 1    &  0.92  &  0.92 \\
			FDP        & 0.13  &  0.19      & 0.88      &  198.5  & 4.59   & 0.91 &  0.68  &  0.68  \\
			die linke  & 0.09  &  0.83      & 0.14      &  1138.5 & 814.5  & 0.001&  0.0007  &  0.0002 \\
			Gruenen   & 0.13   &  0.81      & 0.23      &  378.5  & 237.6  & 0.15 &  0.12  &  0.075  \\
			CSU        & 0.37   &  6.75e-05  & 6.61e-05  &  35.6   & 0.002  & 1    &  0.004  &  0.004 \\
		\end{tabular}
	\end{center} 
	\caption{Results for Germany, only the ``old'' states ,2017. }
	\label{tabReinfGermOld}
\end{table}

\FloatBarrier
\subsection{Details -- France/Le Pen}
The presidential elections in France have (in principle/mostly) two rounds, a first round and runoff elections, in case that no candidate receives more than 50\% of votes in the first round. The elections are direct, each vote counts the same. \\
The data are provided by the France Government. We used data on the level of d\'epartements.\\
\url{https://www.data.gouv.fr/fr/posts/les-donnees-des-elections/}.\par\medskip

{\scriptsize
	\rotatebox{90}{
		\begin{tabular}{lllllllllllll}
			year& candid. & $\hat\nu$ & $\hat\theta$ & $\hat\psi$ & $\hat s$ & $\theta_2$ & $p_{ll}$ & $p_{ks}$ (Reinf.) &  $p_{ks}$ (beta)\\
			\hline
			1965 & MITTERRAND (CIR) &0.29621 &0.00006 &0.00007 &16.70000 &0.00092 &1.00000 &0.00264 &0.00264 \\
			1965 & LECANUET (MRP) &0.14076 &0.20379 &0.00007 &64.50000 &13.14346 &1.00000 &0.90755 &0.91032 \\
			1965 & DE GAULLE (UNR) &0.66489 &0.78879 &0.69318 &873.50000 &211.39934 &0.00000 &0.12579 &0.00091 \\
			1965 & TIXIER-VIGNANCOUR (EXD) &0.03379 &0.79450 &0.00007 &700.50000 &556.51034 &0.00000 &0.00258 &0.00004 \\
			1969 & DUCLOS (PCF) &0.17985 &0.00005 &0.00007 &18.10000 &0.00090 &1.00000 &0.10169 &0.10171 \\
			1969 & DEFFERRE (SFIO) &0.18107 &0.91352 &0.35652 &473.50000 &278.33585 &0.00000 &0.02718 &0.00083 \\
			1969 & POMPIDOU (UDR) &0.87801 &0.76567 &0.99995 &446.50000 &0.01612 &0.00000 &0.25179 &0.00015 \\
			1969 & POHER (CD) &0.22691 &0.00008 &0.00007 &70.00000 &0.00550 &1.00000 &0.12838 &0.12844 \\
			1974 & MITTERRAND (PS) &0.19448 &0.69360 &0.00007 &281.50000 &195.23601 &0.01984 &0.04230 &0.00419 \\
			1974 & GISCARD D'ESTAING (RI) &0.31544 &0.05921 &0.00007 &30.70000 &1.81770 &1.00000 &0.69891 &0.69903 \\
			1974 & CHABAN-DELMAS (UDR) &0.12745 &0.80028 &0.12755 &901.50000 &629.42834 &0.00000 &0.00108 &0.00000 \\
			1974 & ROYER (DVD) &0.02110 &0.80942 &0.00007 &979.50000 &792.77773 &0.00000 &0.00000 &0.00000 \\
			1981 & MARCHAIS (PCF) &0.12149 &0.54736 &0.13981 &62.00000 &29.19165 &0.80061 &0.25322 &0.33004 \\
			1981 & MITTERRAND (PS) &0.28329 &0.16863 &0.99993 &140.50000 &0.00157 &1.00000 &0.17275 &0.17276 \\
			1981 & GISCARD D'ESTAING (UDF) &0.32811 &0.46247 &0.61799 &104.50000 &18.46186 &0.99834 &0.14571 &0.12528 \\
			1981 & CHIRAC (RPR) &0.30222 &0.83611 &0.35581 &678.50000 &365.44584 &0.00000 &0.11781 &0.00019 \\
			1988 & MITTERRAND (PS) &0.33390 &0.00006 &0.00007 &84.00000 &0.00486 &1.00000 &0.24234 &0.24247 \\
			1988 & BARRE (UDF) &0.15714 &0.00004 &0.00007 &95.50000 &0.00416 &1.00000 &0.25615 &0.25586 \\
			1988 & CHIRAC (RPR) &0.31120 &0.82687 &0.36834 &532.50000 &278.12750 &0.00000 &0.02471 &0.00059 \\
			\rowcolor{MyGray}
			1988 & LE PEN (FN) &0.18816 &0.76069 &0.29438 &252.50000 &135.53205 &0.01609 &0.87942 &0.23976 \\
		\end{tabular}
	}
}

\begin{table}
	{\scriptsize 
		\rotatebox{90}{
			\begin{tabular}{lllllllllllll}
				year& candid. & $\hat\nu$ & $\hat\theta$ & $\hat\psi$ & $\hat s$ & $\theta_2$ & $p_{ll}$ & $p_{ks}$ (Reinf.) &  $p_{ks}$ (beta)\\
				\hline
				1995 & JOSPIN (PS) &0.27240 &0.25114 &0.99993 &134.50000 &0.00223 &1.00000 &0.57881 &0.56790 \\
				1995 & BALLADUR (UDF) &0.17659 &0.00005 &0.00007 &85.00000 &0.00395 &1.00000 &0.15455 &0.15375 \\
				1995 & CHIRAC (RPR) &0.29723 &0.82776 &0.33493 &997.50000 &549.13604 &0.00000 &0.06349 &0.00000 \\
				\rowcolor{MyGray}
				1995 & LE PEN (FN) &0.13366 &0.16371 &0.00007 &57.50000 &9.41255 &1.00000 &0.35950 &0.35765 \\
				2002 & JOSPIN (PS) &0.25701 &0.83780 &0.31043 &890.50000 &514.45847 &0.00000 &0.59056 &0.00649 \\
				2002 & BAYROU (UDF) &0.05982 &0.00006 &0.00007 &114.50000 &0.00740 &0.99922 &0.00000 &0.00000 \\
				2002 & CHIRAC (UMP) &0.27697 &0.82723 &0.31351 &1069.50000 &607.35573 &0.00000 &0.03866 &0.00000 \\
				\rowcolor{MyGray}
				2002 & LE PEN (FN) &0.14604 &0.00004 &0.00007 &35.30000 &0.00149 &1.00000 &0.06609 &0.06608 \\
				2007 & ROYAL (PS) &0.33165 &0.77302 &0.38618 &419.50000 &199.05094 &0.00014 &0.13556 &0.03850 \\
				2007 & BAYROU (UDF) &0.17800 &0.00005 &0.00007 &108.50000 &0.00504 &1.00000 &0.06028 &0.06028 \\
				2007 & SARKOZY (UMP) &0.48899 &0.81562 &0.50867 &1135.50000 &455.03893 &0.00000 &0.41382 &0.00342 \\
				\rowcolor{MyGray}
				2007 & LE PEN (FN) &0.08921 &0.00004 &0.00007 &54.00000 &0.00235 &1.00000 &0.11541 &0.11533 \\
				2012 & Jean-Luc MELENCHON (FG) &0.10230 &0.00004 &0.00007 &93.00000 &0.00380 &1.00000 &0.00051 &0.00050 \\
				2012 & François HOLLANDE (PS) &0.41777 &0.80629 &0.45421 &599.50000 &263.81758 &0.00000 &0.28500 &0.00830 \\
				2012 & Nicolas SARKOZY (UMP) &0.40273 &0.81085 &0.44187 &610.50000 &276.28720 &0.00000 &0.29963 &0.00091 \\
				\rowcolor{MyGray}
				2012 & Marine LE PEN (FN) &0.15076 &0.00005 &0.00007 &28.60000 &0.00138 &1.00000 &0.01465 &0.01466 \\
				\rowcolor{MyGray}
				2017 & LE PEN &0.19953 &0.00006 &0.00007 &23.40000 &0.00140 &1.00000 &0.05362 &0.05371 \\
				2017 & MLENCHON &0.23086 &0.47868 &0.53214 &104.50000 &23.40335 &1.00000 &0.02407 &0.01695 \\
				2017 & MACRON &0.28845 &0.39413 &0.80474 &112.50000 &8.65774 &0.31516 &0.10480 &0.03947 \\
				2017 & FILLON &0.29702 &0.82423 &0.35947 &457.50000 &241.53438 &0.00000 &0.07550 &0.00000 \\
			\end{tabular}
	}}
	\caption{Results for France - departement level.}
	\label{tabReinfFrance}
\end{table}

\newpage
The estimates for Le Pen from 2017 indicate that the model finds no reinforcement aspects in the data of Le Pen ($\hat\theta\approx 0$). Accordingly, the LL-test indicates that the zealot model performs as well as the reinforcement-model. The KS-test indicates that the data (canton level) do not follow the distribution predicted by the reinforcement model, while the data on district level are close to th reinforcement model (cannot rejected at the significance level of $0.01$, but only at a significance level of $0.05$).
Note that the number of data on canton level ($n=35703$) is much larger than that on  departments level ($n= 2090$), which explains that even small deviations from the model distribution leads to extremely significant values in the KS-test. 
\begin{table}[h!]
	\begin{center}
		\begin{adjustbox}{width=\columnwidth,center}
			\begin{tabular}{l|llll|l|lll}
				election(2017) & $\hat\nu$ & $\hat\theta$ & $\hat\psi$ & $\hat s$ & $\theta_2$ & $p_{ll}$ & $p_{ks}$ (Reinf.) &  $p_{ks}$ (beta)\\
				\hline
				first round/canton  &0.23063 &0.00005 &0.00007 &15.90000 &0.00074 &1.00000 &0.00000 &0.00000 \\
				first round/departement  &
				0.19953 &0.00006 &0.00007 &23.40000 &0.00140 &1.00000 &0.05362 &0.05371 \\
				second round/canton  &0.39724 &0.05783 &0.00007 &12.40000 &0.71699 &1.00000 &0.00000 &0.00000 \\
			\end{tabular}
		\end{adjustbox}
	\end{center}
	\caption{Estimates for Le Pen in the presidential elections (2017), different data sets: First round and second round (canton level), first round (departement level). }\label{LePenEstim}
\end{table}

\FloatBarrier

\subsection{Details -- The Netherlands/ Catholic People's Party}
The parliament's election in The Netherlands do not have a threshold, but are purely proportional. 
That's interesting as the effect of strategic voting will be less prominent as e.g.\ in Germany, where
a 5\% threshold is implemented. The election districts, however, have a very different size as only the 
total number of votes, all over the country, counts. That might disturb our assumption that all election 
districts are i.i.d. \\
The Catholic People's party did play a central role after the second world war. After 1971, it did lose 
importance, and eventually merged with other parties.\\
The model indicates that the success of the party is almost exclusively due to reinforcement ($\hat \nu\approx 0$), which explains the peak of the distribution at a vote share of zero (Fig.~2 in the main paper).\\
The data can be found at the internet-pages of the Dutch government,\\
\url{https://www.verkiezingsuitslagen.nl/verkiezingen} 

\begin{table}
	\begin{center}
		\begin{tabular}{ll|llll|l|lll}
			year&	party & $\hat\nu$ & $\hat\theta$ & $\hat\psi$ & $\hat s$ & $\theta_2$ & $p_{ll}$ & $p_{ks}$ (Reinf.) &  $p_{ks}$ (beta)\\
			\hline
			1946  &   (KVP)  & 6.61e-05  &  0.98   & 0.46   &  29.9 &15.9  & $<$1e-10  &   $<$1e-10  &  0 \\
			1948  &   (KVP)  & 6.61e-05  &  0.98  & 0.44    &  31.0 &16.8  &  $<$1e-10  &   $<$1e-10  &  0 \\
			1952  &   (KVP)  & 6.61e-05  &  0.98  & 0.39    &  34.0 &19.7  &  $<$1e-10  &   $<$1e-10  &  0 \\
			1956  &   (KVP)  & 6.61e-05  &  0.98  & 0.44    &  30.2 &16.5  &  $<$1e-10 &   $<$1e-10 &  0 \\
			1959  &   (KVP)  & 6.61e-05  &  0.96  & 0.39    &  35.4 &20.7  &  $<$1e-10  &  $<$1e-10  &  0 \\
			1963  &   (KVP)  & 6.61e-05  &  0.93  & 0.32    &  40.9 &26.1  &  $<$1e-10  &  3.21e-06  &  0 \\
			1967  &   (KVP)  & 6.61e-05  &  0.86  & 0.10    &  71   &54.5  &  $<$1e-10  &  5.1e-05  &  0 \\
			1971  &   (KVP)  & 6.61e-05  &  0.82  & 6.61e-05&  95   &77.5  &  $<$1e-10  &  0.0003  &  0 \\
			1972  &   (KVP)  & 6.61e-05  &  0.80  & 6.61e-05&  65   &52.1  &  $<$1e-10  &  1.26e-05  &  0 \\
		\end{tabular}
	\end{center} 
	\caption{Results for NL/KVP. We find a transition from bimodal to unimodal during the years. }
	\label{tabReinfGermOld}
\end{table}

\end{appendix}
\end{document}